\documentclass[aps,showpacs,amsmath,amssymb,superscriptaddress,nofootinbib,twocolumn]{revtex4}
\usepackage{multirow}
\usepackage{bigstrut}
\usepackage{amssymb}
\usepackage{amsmath}
\usepackage{graphicx}
\usepackage{dcolumn}
\usepackage{bm}
\usepackage{CJK}
\usepackage{relsize}
\usepackage{cancel}
\usepackage{color}

\begin{document}

\title{Phonon-induced Majorana Qubit Relaxation in Tunnel-Coupled Two-Island Topological Superconductors}

\author{Yang Song}\email{ysong128@umd.edu}\affiliation{Condensed Matter Theory Center and Joint Quantum Institute, Department of Physics, University of Maryland, College Park, Maryland 20742}
\author{S.~Das Sarma}
\affiliation{Condensed Matter Theory Center and Joint Quantum Institute, Department of Physics, University of Maryland, College Park, Maryland 20742}

\begin{abstract}
A Majorana-based qubit can form within the fermion parity subspace of a two-island topological superconductor setup consisting of four Majorana zero modes.   We study the relaxation time of this Majorana qubit induced by the intrinsic Majorana-phonon interaction in combination with single electron tunneling between the two islands. The theory is developed for  both the spinless $p$-wave Kitaev chain model and the spin-orbit-coupled semiconductor nanowire with induced superconductivity and Zeeman splitting. We systematically discuss the dependence of the relaxation rate on Majorana charge distribution (varying with the chemical potential, magnetic field, superconducting pairing, spin-orbit coupling and island length) as well as on phonon wavelength and the length of the tunnel barrier. Importantly, the accompanied  phonon energy can be tuned by the voltage bias over the tunnel barrier, leading to relatively easy manipulation and suppression of the relaxation rate.
\end{abstract}
\maketitle

\section{Introduction}\label{sec:outline}

Quantum computing promises to explore and exploit intrinsic quantum mechanical principles such as superposition \cite{Nielsen_Chuang_book}, producing algorithms that dramatically enhance the performance of certain computational tasks, such as prime factorization and solving intractable quantum many body problems \cite{Preskill_arxiv12}. However,  severe delicacy of the quantum state comes with this superposition principle, and greatly increases the chance of computational errors compared to classical digital computing. To address this outstanding challenge, fault-tolerant quantum error correction has been proposed provided that the errors are small \cite{Devitt_RPP13, Gottesman_PSAM10}. In practice, such quantum error correction implementation requires long coherence time for the qubit as well as necessitating a huge number of qubits, presenting enormous technical difficulties. A different and ingenious approach to quantum computing avoids quantum error correction complications, instead using the physical error protection provided by topological properties of the suitably chosen qubits \cite{Kitaev_AP03, Nayak_RMP08}. Several potential topological qubit candidates  are the fractional quantum Hall system \cite{Moore_NucPhys91}, the $p$-wave superconductor \cite{DasSarma_PRB06}, and the topological superconductor (TSC) by proximitizing an $s$-wave superconductor to a topological insulator \cite{Fu_PRL08, Fu_PRB09} or a semiconductor with large spin-orbit coupling \cite{Sau_PRL10, Lutchyn_PRL10, Oreg_PRL10}. The most accessible and experimentally studied material system for topological qubits is the strongly spin-orbit-coupled one-dimensional (1D) semiconductor nanowire with induced superconductivity and magnetic field-induced Zeeman splitting, where one Majorana zero mode (MZM) is located at each end of the wire \cite{Kitaev_PhysUsp01, Lutchyn_PRL10, Oreg_PRL10}. This MZM-based hybrid superconductor-semiconductor nanowire system has been extensively studied both experimentally and theoretically over the last 10 years in the context of developing a topological qubits \cite{Alicea_RPP12, Leijinse_SST12, Stanescu_JPCM13, Beenakker_ARCMP13, Elliott_RMP15, DasSarma_QI15, Beenakker_RMP15, Sato_JPSJ16, Aguado_NC17, Lutchyn_arxiv17}.

While two MZMs form a pair of quantum degenerate many-body fermion states, i.e., occupied and unoccupied, the transition between the two must rely on an external supply of fermions. In fact, the minimal Majorana-based topological qubit consists of four MZMs in a two-island setup with each island supporting two MZMs localized at its ends.   The qubit is defined in the same total fermion parity subspace of the four MZMs. Typical gate operations involve careful braiding of these MZMs around each other to achieve desired outcomes. To perform topologically protected gate operations, additional structures are utilized such as T and Y junctions \cite{Alicea_NatPhys11, Aasen_PRX16}. Alternative schemes such as Majorana box qubits and measurement-only protocols without braiding operations are also proposed \cite{Plugge_NJP17, Karzig_PRB17}. The tunnel-coupled TSC islands consisting of four MZMs are the ubiquitous building blocks in all these designs. Experimentally, with the steady improvement in device fabrication and experimental measurement \cite{Mourik_Science12, Rokhinson_NatPhys12, Deng_NanoLett12, Churchill_PRB13, Das_NatPhys12, Finck_PRL13, Albrecht_Nature16, Deng_Science17, Zhang_Nature18}, the two-island setup is naturally the next step to demonstrate Majorana qubit  operations (possibly non-topological) as well as the Rabi oscillation  \cite{Bauer_arxiv18}. We focus on this two-island and four-MZM setup (a schematic shown in Fig.~\ref{fig_schematic}) in our study.

In this minimal two-island four-MZM platform, however, whereas the fermion parity-conserved Majorana qubit facilitates qubit operations, it also opens up ways for qubit transitions  through inter-island electron tunneling, which are not protected against by either topology or fermion parity. These operations can be intentional as part of the universal quantum gate operations \cite{Schmidt_PRL13}, or unintentional (arising from ``errors''). When combined with the right compensation energy for  the tunneling electron, the qubit relaxation happens. The energy quanta may be provided by intrinsic phonons, environmental charge noise, etc. In this early development stage of experimentally implementing two-island MZMs, our paper  focuses on the Majorana qubit relaxation induced by the intrinsic Majorana-phonon interaction and inter-island tunneling. Since electron-phonon interaction is intrinsic, it is always present as a possible error source without any topological protection in the two-island tunneling set up, and as such, understanding the phonon-induced Majorana qubit relaxation is of crucial importance. Some of the physics studied here has been touched upon recently also by Knapp \textit{et al} \cite{Knapp_PRB18}, where they focused on the Majorana dephasing of a single island. Our focus is on the two-island experimental qubit geometry. We systematically map out the dependence of two-island qubit relaxation on all key physical parameters. We discuss the detailed dependence on chemical potential ($\mu$), Zeeman energy ($V_z$), superconducting pairing potential ($\Delta$), spin-orbit coupling ($\alpha$) and the island length ($l$) as well as on the phonon wavelength ($2\pi/q_0$) and the length of the tunnel barrier ($l_B$). The coherence time for a Majorana box qubit where the noise environment is modeled as a single external capacitor was studied in Ref.~\cite{Munk-Nielsen_thesis17}. We emphasize that our work is different from more generic earlier works on the coherence of the MZM \cite{Goldstein_PRB11, Cheng_PRB11, Cheng_PRB12,Budich_PRB12, Rainis_PRB12, Schmidt_PRB12, Ho_NJP14, Hu_PRB15, Ippoliti_PRA16} whose results  apply in general to the MZM in each of our two islands. Our work looks into the additional features brought about by the specific tunnel coupled two-island setup, and we do not consider above-gap quasiparticle states, assuming superconducting leads and low temperatures. We are specifically interested in the phonon-induced inter-island tunneling relaxation of the Majorana qubit---thus, MZM, phonon, and tunneling are all important and essential ingredients of the physics we are interested in.

It is crucial to understand the effect of phonon-assisted tunneling ($\tilde{T}$) in this Majorana qubit.   $\tilde{T}$ term acts as a $\sigma_x$ operator in the qubit space, whose basis states are $|0\rangle(|1\rangle)=$ Majorana states are occupied (unoccupied) in both islands, or   $|0\rangle(|1\rangle)=$ the Majorana state is occupied and unoccupied respectively in the left (right) and right (left) island.   It is the new term which must be considered in the two-island setup. This physics does not exist in the one-island MZM geometry. As mentioned, this tunneling relaxation is much more efficient than the Majorana relaxation in the one-island case, since it is not prevented by fermion parity conservation and since the energy compensation in this relaxation
is often much smaller than the TSC pairing $\Delta$. This relaxation is also strongly enhanced for a short island where larger MZM charge responds more effectively to the electric field perturbation (including electron-phonon interaction or charge noise), a feature mostly overlooked in the studies so far. In this sense, this Majorana-based qubit is somewhat similar to the double-quantum-dot (DQD) charge qubit \cite{Hayashi_PRL03, Petta_PRL04}, but with a tunable charge less than that of an electron. In fact, the effective charge goes to zero asymptotically as the MZMs are infinitely separated from each other, becoming isolated Majorana modes in the well-separated limit. It has been known that spontaneous phonon emission plays an evident role in the DQD relaxation \cite{Fujisawa_science98, Hayashi_PRL03}, and a similar relaxation phenomenon applies in the two-island Majorana case also.  A noisy $\tilde{T}(t)$ term may have two practical effects on the Majorana qubit: (1) A gate error for a $\sigma_x$ gate rotation, including any possible Rabi oscillation demonstration of the Majorana qubit. This error usually only operates for a short time period, but the  noise amplitude may be large, due to the intentional lowering of the tunnel barrier during the active gate operation leading to strong inter-island wavefunction overlap. (2) Qubit flip without active gate operations, i.e., waiting between gate operations or serving as a qubit memory. The exposure time to errors  is long in this case, though the  noise amplitude can be made relatively small by intentionally ``turning off'' the tunneling (but not completely off). They are analogous  to the situations in the ordinary semiconductor DQD qubit \cite{Hu_PRL06} with each island here acting as an effective QD.

The rest of this paper is organized as follows. We start with a general Majorana-phonon interaction matrix element in Sec.~\ref{sec:A}, and combine it with the tunnel coupling between two TSC islands to derive the phonon-assisted qubit relaxation rate in Sec.~\ref{sec:B}. We simplify the relaxation rate theory into a 1D formulation  and estimate the numerical prefactor for some typical materials parameters. Note that since we use a second-order perturbation theory to get the phonon-assisted tunnel coupling $T$, the result applies when  $T$ is much smaller than the phonon energy $\varepsilon_{q_0}$. We leave $T$ and $\varepsilon_{q_0}$ as (unknown) experimental variables, and focus on evaluating the core unitless integral $\Gamma$ for the relaxation rate [Eq.~(\ref{eq:Gamma})]. Obtaining and evaluating the dimensionless expression for $\Gamma$ for the phonon-assisted two-island Majorana tunneling is our main goal in the current work. We respectively apply the theory to the Kiteav chain model in Sec.~\ref{sec:Kitaev_chain} and the superconducting spin-orbit nanowire model in Sec.~\ref{sec:semiconductor_model}. Our reason for showing detailed numerical results for the Kitaev model (Sec.~\ref{sec:Kitaev_chain}) is that many features of the relaxation phenomenon in realistic nanowires (Sec.~\ref{sec:semiconductor_model}) become apparent already in the simpler Kitaev model which has many fewer parameters than the nanowire case. The numerical calculation uses 1D tight-binding model, and the representative results and trends are shown in Figs.~\ref{fig_mu0}-\ref{fig_Gamma_N100_q0l1-3-10}. These direct results are compared  with analytical estimations whenever possible. We conclude in Sec.~\ref{sec:summary} with a summary and discussion,  and present some more detailed numerical results for $\Gamma$ in the Appendix.

\section{Majorana-phonon matrix element within one island}\label{sec:A}

Within the harmonic approximation, the general electron-phonon (e-ph) interaction Hamiltonian is
\begin{eqnarray}
H_{\rm ep} = \sum_i c^\dag_i c_i \sum_{\mathbf{q},\lambda} (\alpha_{\mathbf{q},\lambda,i} a^\dag_{\mathbf{q},\lambda} + \alpha^*_{\mathbf{q},\lambda,i} a_{\mathbf{q},\lambda}) ,
\label{eq:H_ep}
\end{eqnarray}
where $i$ goes over lattice sites $\mathbf{R}_i$  and also over spins except for the spinless Kitaev chain model, $\mathbf{q}$ and $\lambda$ are the wavevector and branch of the phonon, $c$ and $a$ are the conduction-band electron and phonon operators, and the e-ph coupling
\begin{eqnarray}
\alpha_{\mathbf{q},\lambda,i} = (i\bm\Xi + \frac{e_0\mathbf{q}\cdot \mathbf{e}_m}{q^2 \epsilon})\sqrt{\frac{\hbar}{2 V \rho \omega_{\mathbf{q},\lambda}}} \mathbf{q}\hat{\mathbf{e}}_\lambda e^{i \mathbf{q}\cdot \mathbf{R}_i},
\label{eq:alpha_ep}
\end{eqnarray}
where $\bm\Xi$ and $\mathbf{e}_m$ are the deformation-potential tensor  and the piezoelectric tensor,  $e_0$ and $\epsilon$ are  the electron charge and dielectric constant, $\omega_{\mathbf{q},\lambda}$ and $\hat{\mathbf{e}}_\lambda$ are the phonon energy and polarization direction, respectively, and $V$ and $\rho$ are the material volume and density. These parameters measure the e-ph coupling strength in a particular material, which may depend on the position in a semiconductor-superconductor hybrid structure. $\alpha_{\mathbf{q}} =\alpha^*_{-\mathbf{q}}$. Here only low-energy acoustic phonons are considered. It is unlikely that optical phonons of the semiconductor are relevant for Majorana relaxation.

For e-ph interaction within the same island, we are interested in two possible matrix elements involving the ground states of the topological superconductor, i.e., the MZMs,
\begin{eqnarray}\label{eq:M_e}
M^{\rm e}_{{\rm ep},\mathbf{q}}=\langle N_{\mathbf{q},\lambda}=1| \langle 0_b|H_{\rm ep} |0_b\rangle |N_{\mathbf{q},\lambda}=0 \rangle ,
\end{eqnarray}
where e denotes even fermion parity, and
\begin{eqnarray}\label{eq:M_o}
M^{\rm o}_{{\rm ep},\mathbf{q}}=\langle N_{\mathbf{q},\lambda}=1| \langle 0_b|b_0 H_{\rm ep} b^\dag_0|0_b\rangle |N_{\mathbf{q},\lambda}=0 \rangle ,
\end{eqnarray}
where o denotes odd fermion parity. $|0_b\rangle$ is defined such that $b_l|0_b\rangle =0$, where the quasiparticle operator $b_0$ is for the Majoranan mode and $b_k$ ($k\neq 0$) are for the bulk quasiparticle modes. $|N_{\mathbf{q},\lambda} \rangle$ is the phonon state with associated occupation number $N_{\mathbf{q},\lambda}$. We make the reasonable assumption that all phonon absorption processes are exponentially suppressed by having a low operational temperature and only consider phonon emission process [otherwise one simply adds the Bose-Einstein factor $1/(e^{\hbar \omega_q/k_BT}-1)$ to the number of emitted as well as the absorbed phonons]. We do not consider the transition between MZMs and bulk quasiparticles or between two bulk quasiparticles. We are thus studying the minimal theoretical model for phonon relaxation, assuming only phonon emission and neglecting all quasiparticle poisoning.

Substituting $H_{\rm ep}$ of Eq.~(\ref{eq:H_ep}) into Eq.~(\ref{eq:M_e}), we have
\begin{eqnarray}
M^{\rm e}_{{\rm ep},\mathbf{q}}
= \sum_i  \alpha_{\mathbf{q},\lambda,i} \langle 0_b|  c^\dag_i c_i  |0_b\rangle.
\end{eqnarray}
 
 One solves $\langle 0_b|  c^\dag_i c_i  |0_b\rangle$ by expanding $c_i$ into the eigenmodes of the system.
\begin{eqnarray}
&&\langle 0_b|  c^\dag_i c_i  |0_b\rangle
\nonumber\\
&=&  \langle 0_b| \sum_{l=0,k} (u_{i,l}b_l+ v^*_{i,l}b^\dag_l)  \sum_{l'=0,k} (v_{i,l'}b_{l'}+ u^*_{i,l'}b^\dag_{l'})   |0_b\rangle
\nonumber\\
&=& \sum_{l=0,k} |u_{i,l}|^2 \langle 0_b| b_l b^\dag_l  |0_b\rangle = \sum_{l=0,k} |u_{i,l}|^2,  \label{eq:even_occupation}
\end{eqnarray}
where $b^\dag_l= \sum_i (u_{i,l}c^\dag_i+ v_{i,l} c_i)$.  This quantity depends on the MZM as well as the bulk quasiparticle ground states.  Physically, it is the occupation number expectation at site $i$ for state $|0_b\rangle$. In principle, one could diagonalize the Bogoliubov–de Gennes (BdG) Hamiltonian matrix and obtain all $N$ (number of sites) eigenvectors. Similarly, for the e-ph coupling between the $b^\dag_0|0_b\rangle$ state, we have
\begin{eqnarray}
&&\langle 0_b| b_0  c^\dag_i c_i  b^\dag_0|0_b\rangle
= \sum_{l=k} |u_{i,l}|^2 +|v_{i,0}|^2
\nonumber\\
&=& \langle 0_b|  c^\dag_i c_i  |0_b\rangle- |u_{i,0}|^2 +|v_{i,0}|^2 . \label{eq:odd_occupation}
\end{eqnarray}
They show that the charge density of $b^\dag_0|0_b\rangle$ and $|0_b\rangle$ states differ by the net charge density of the MZM which makes physical sense.

\section{Phonon-assisted inter-island tunneling}\label{sec:B}

The Majorana-phonon interaction matrix elements derived above, together with tunneling across the two islands, induce the phonon-assisted tunneling of Majorana mode and the switch of Majorana occupation in each island. Note that the intra-island Majorana-phonon interaction alone cannot switch the Majorana mode as it conserves fermion number within the island. The Majorana qubit transition, as defined below, requires the presence of the inter-island tunnel coupling.

The Majorana qubit can be defined within the even or odd total fermion parity space. To be concrete, let us assume that the qubit is made of states $|0_Q\rangle \equiv | {\rm e}_L\rangle\otimes | {\rm o}_R\rangle$, and $|1_Q\rangle \equiv | {\rm o}_L\rangle\otimes | {\rm e}_R\rangle$,  where $L$ or $R$ labels the left or right island and $|{\rm e}\rangle\equiv |0_b\rangle$ and $|{\rm o}\rangle\equiv b_0^\dag|0_b\rangle$. The energy $\varepsilon_{0_Q} = \varepsilon_{{\rm e}_L} +\varepsilon_{{\rm o}_R}$,  $\varepsilon_{1_Q} = \varepsilon_{{\rm }o_L} +\varepsilon_{{\rm e}_R}$, and the nominal qubit splitting $\varepsilon_{Q} = \varepsilon_{1_Q}- \varepsilon_{0_Q}$. However, an important aspect of the phonon-induced electron tunneling between the two islands is that, taking $|1_Q\rangle \rightarrow |0_Q\rangle$ for example, the transfer of energy between electronic and phonon systems is
\begin{eqnarray}
\varepsilon_{q_0}=\varepsilon_{Q} \mp (eV^{\rm lead}_L-eV^{\rm lead}_R), \label{eq:E_q0}
 \end{eqnarray}
\textit{but not} merely $\varepsilon_{Q}$, if an electron goes from the left/right island to the right/left island. $V^{\rm lead}_{L/R}$ is the external voltage on two sides of the barrier. This reflects the fact that the difference of the zero energy references in the two TSCs has to be taken into account. (We caution that this potential energy difference between left and right leads may neither be experimentally measurable nor experimentally controllable in a precise manner.)

Putting together all relevant terms including Majorana modes, phonons, their coupling, as well as tunnel coupling, the relevant Hamiltonian  reads as
\begin{eqnarray}
H\!\! &=&\!\! \!\sum_{\nu=0,1} \!\left[\varepsilon_{\nu_Q} |\nu_Q\rangle \langle \nu_Q| \!+\! (\varepsilon_{\nu_Q} \!+\!\hbar\omega_{\mathbf{q},\lambda})|\nu_Q\rangle| 1_{\mathbf{q},\lambda} \rangle \langle \nu_Q|  \langle 1_{\mathbf{q},\lambda} |\right]
\nonumber\\
&& +  (M^{{\rm o}_L}_{{\rm ep},\mathbf{q}}+M^{{\rm e}_R}_{{\rm ep},\mathbf{q}} ) |1_Q\rangle | 1_{\mathbf{q},\lambda} \rangle \langle 1_Q|
 \nonumber\\
 && +  (M^{{\rm o}_R}_{{\rm ep},\mathbf{q}}+M^{{\rm e}_L}_{{\rm ep},\mathbf{q}} ) |0_Q\rangle | 1_{\mathbf{q},\lambda} \rangle \langle 0_Q|  + \textrm{H. c.}
  \nonumber\\
&& + T |0_Q\rangle \langle 1_Q| + T |0_Q\rangle | 1_{\mathbf{q},\lambda} \rangle \langle 1_Q|  \langle 1_{\mathbf{q},\lambda} | + \textrm{H. c.},
\end{eqnarray}
where H.c. denotes Hermitian conjugate, for simplicity $| 1_{\mathbf{q},\lambda} \rangle\equiv | N_{\mathbf{q},\lambda}=1 \rangle$ and we omit $| N_{\mathbf{q},\lambda}=0 \rangle$,  $T$ is the tunneling matrix element between $|1_Q\rangle$ and $|0_Q\rangle$. For $T\ll \hbar \omega_{\mathbf{q},\lambda}$ \cite{footnote}, we can use a second-order perturbation or unitary transformation to fold the e-ph interaction into the qubit subspace, and have a renormalized phonon-assisted tunneling matrix element (see, e.g., \cite{Glazman_JETP88,Brandes_PRB02}),
\begin{eqnarray}
(T  |0_Q\rangle \langle 1_Q|- T^*  |1_Q\rangle \langle 0_Q|) (\kappa a^\dag_{\mathbf{q},\lambda} -\kappa^*a_{\mathbf{q},\lambda}), \label{eq:ph-assisted_tunnel}
\end{eqnarray}
where
\begin{eqnarray}
\kappa_{\mathbf{q}}
& =&(M^{\rm o_R}_{\rm ep,\mathbf{q}}+M^{\rm e_L}_{\rm ep,\mathbf{q}}-M^{\rm e_R}_{\rm ep,\mathbf{q}}-M^{\rm o_L}_{\rm ep,\mathbf{q}})/\hbar\omega_{\mathbf{q},\lambda}
  \nonumber\\
&=& \frac{1}{\hbar\omega_{\mathbf{q},\lambda} }\left[\sum_{i\in L} \alpha_{\mathbf{q},\lambda,i} (|u_{i,0}|^2 -|v_{i,0}|^2 ) \right.
  \nonumber\\
  && \qquad\quad\left.-\sum_{i\in R} \alpha_{\mathbf{q},\lambda,i} (|u_{i,0}|^2 -|v_{i,0}|^2 ) \right]. \label{eq:kappa_q}
\end{eqnarray}
As a result, only contribution of the Majorana component remains and all other quasiparticle components from Eqs.~(\ref{eq:even_occupation}) and (\ref{eq:odd_occupation}) cancel out.   Since $\kappa_{\mathbf{q}} =\kappa^*_{-\mathbf{q}}$,  we have $\sum_{\mathbf{q}} (\kappa_{\mathbf{q}}  a^\dag_{\mathbf{q},\lambda} -\kappa^*_{\mathbf{q}} a_{\mathbf{q},\lambda} ) = \sum_{\mathbf{q}}\kappa_{\mathbf{q}} (a^\dag_{\mathbf{q},\lambda} - a_{-\mathbf{q},\lambda} )$.  There is also a connection between, say, $M^{{\rm o}_L}_{{\rm ep},\mathbf{q}}$ and $M^{{\rm o}_R}_{{\rm ep},\mathbf{q}}$, for two identical islands, by the phonon displacement in the tunnel barrier region  (see a simpler example in Ref.~\cite{Brandes_PRL99}). With this effective dynamical ($\sim e^{i\omega_{\mathbf{q},\lambda} t}$) tunnel coupling in Eq.~(\ref{eq:ph-assisted_tunnel}), we can next calculate the relaxation rate between $|1_Q\rangle$ and $|0_Q\rangle$ states.

With Eqs.~(\ref{eq:ph-assisted_tunnel}), (\ref{eq:kappa_q}) and (\ref{eq:alpha_ep}),  the relaxation rate from $|1_Q\rangle$ to $|0_Q\rangle$ (assuming $\varepsilon_{q_0}>0$) at low temperature limit is (within the leading-order time-dependent perturbation theory)
\begin{eqnarray}
\tau^{-1}\!&\! =\! &\!\frac{|T  |^2}{2(2\pi)^2 \hbar^2 \rho }  \sum_\lambda  \int d^3\mathbf{q}  \frac{|(i\bm\Xi + \frac{e_0\mathbf{q}\cdot \mathbf{e}_m}{q^2 \epsilon}) \mathbf{q}\cdot\hat{\mathbf{e}}_\lambda|^2}{ \omega^3_{\mathbf{q},\lambda}}
 \nonumber\\
&&\!\!\!\! \left|\sum_{i\in L} e^{i \mathbf{q}\cdot \mathbf{R}_i} (|u_{i,0}|^2 -|v_{i,0}|^2 ) \right.
 \nonumber\\
&&\left.-\sum_{i\in R} e^{i \mathbf{q}\cdot \mathbf{R}_i}  (|u_{i,0}|^2 -|v_{i,0}|^2 ) \right|^2
 \delta(\hbar \omega_{\mathbf{q},\lambda}- \varepsilon_{q_0} ).\qquad \label{eq:rate_general}
\end{eqnarray}

In the one-dimensional limit, only two phonon states of wavevector $\pm q_0$  satisfy the energy requirement, $\hbar v_l q_0= \varepsilon_{q_0}$, with $v_l$ the longitudinal phonon velocity and $\hat{\mathbf{e}}_\lambda$ aligned along the wire direction.  The relaxation rate reduces to a simple algebraic expression in the 1D limit:
\begin{eqnarray}
\tau^{-1}_{1D} =
 \frac{(|\Xi_l|^2 + |\frac{e_0 e_{m,l}\hbar v_l}{\varepsilon_{q_0} \epsilon}|^2)|T |^2 }{\hbar^2  v^3_{l} \rho_l \varepsilon_{q_0}}   \Gamma,
\label{eq:rate_1D}
\end{eqnarray}
where
\begin{eqnarray}
\Gamma &\equiv& \left|\sum_{i\in L} e^{ i  \varepsilon_{q_0} R_i/\hbar v_l} (|u_{i,0}|^2 -|v_{i,0}|^2 )
\right.
 \nonumber\\
&&\left.
\qquad-\sum_{i\in R} e^{ i  \varepsilon_{q_0} R_i/\hbar v_l}  (|u_{i,0}|^2 -|v_{i,0}|^2 ) \right|^2, \label{eq:Gamma}
\end{eqnarray}
and $\rho_l$ is the linear density of the wire, $\Xi_l$ and $e_{m,l}$  are the deformation potential and piezoelectric constant for the longitudinal mode, respectively. Apparently this result requires $\varepsilon_{q_0}$ to be smaller than the linear acoustic phonon region, the order of 10 meV, which should be trivially satisfied for the Majorana qubit to have any usefulness at all. We do not include any interference between the deformation potential and piezoelectric channels, which is small (and higher-order) any way.

In the following two sections, we study the core quantity of unitless $\Gamma$ in the Kitaev chain and the superconducting spin-orbit nanowire models, respectively, with the prefactor to be specified later for given materials. Once $\Gamma$ is known, the calculation of the phonon-induced relaxation as a function of inter-island tunneling becomes straightforward using Eq.~(\ref{eq:rate_1D}).

\section{Kitaev chain model}\label{sec:Kitaev_chain}

For the spinless p-wave Kitaev chain, one has the analytical solution for $u_{i,0}$ and $v_{i,0}$ in certain parameter regimes. To be specific, the tight-binding Hamiltonian we use is
\begin{eqnarray}
H_{\rm pw} =-\mu c^\dag_i c_i - t (c^\dag_i c_{i+1} + H.c.) -\Delta (c^\dag_i c^\dag_{i+1} + H.c.),\;\;
\end{eqnarray}
where $\mu, t$, and $\Delta$ are all real. The corresponding BdG Hamiltonian is,
\begin{eqnarray}
H^{\rm BdG}_{\rm pw} \!\!&=&\! \frac{1}{2}\left\{\sum_{i}
 (c^\dag_i\; c_i) \left( \begin{array}{cc} -\mu & 0 \\ 0 & \mu \end{array} \right) \left( \begin{array}{c} c_i \\ c^\dag_i  \end{array} \right)
 \right.
 \nonumber\\
&& \quad +
\left.
\left[(c^\dag_i\; c_i) \left( \begin{array}{cc} -t \!& \!-\Delta \\ \Delta \!&\! t \end{array} \right)\! \left( \begin{array}{c} c_{i+1} \\ c^\dag_{i+1}  \end{array} \right)\! +\! H.c. \right] \!\right\}.\quad
\end{eqnarray}
The eigenvectors of the matrix represent the Bogoliubov quasiparticle $b^\dag_l= \sum_i u_{i,l}c^\dag_i+ v_{i,l} c_i$ associated with the eigenvalue $\varepsilon_l\ge 0$ and its partner $b_l= \sum_i v^*_{i,l}c^\dag_i+ u^*_{i,l} c_i$ with the eigenvalue $-\varepsilon_l$, except that in the case of doubly degenerate eigenvectors at $\varepsilon_l= 0$ one needs to impose the above linear combination and assigns either vector to $b^\dag_0$ and the other to $b_0$. In the $\varepsilon_l= 0$ case, one can also obtain two vectors $\gamma_{1,2}$ such that $\gamma_{1,2}^\dag= \gamma_{1,2}$ [for example, by $b^\dag_0+b_0$ and $i(b^\dag_0-b_0)$].

\begin{figure}[!htbp]
\includegraphics[width=8.5cm]{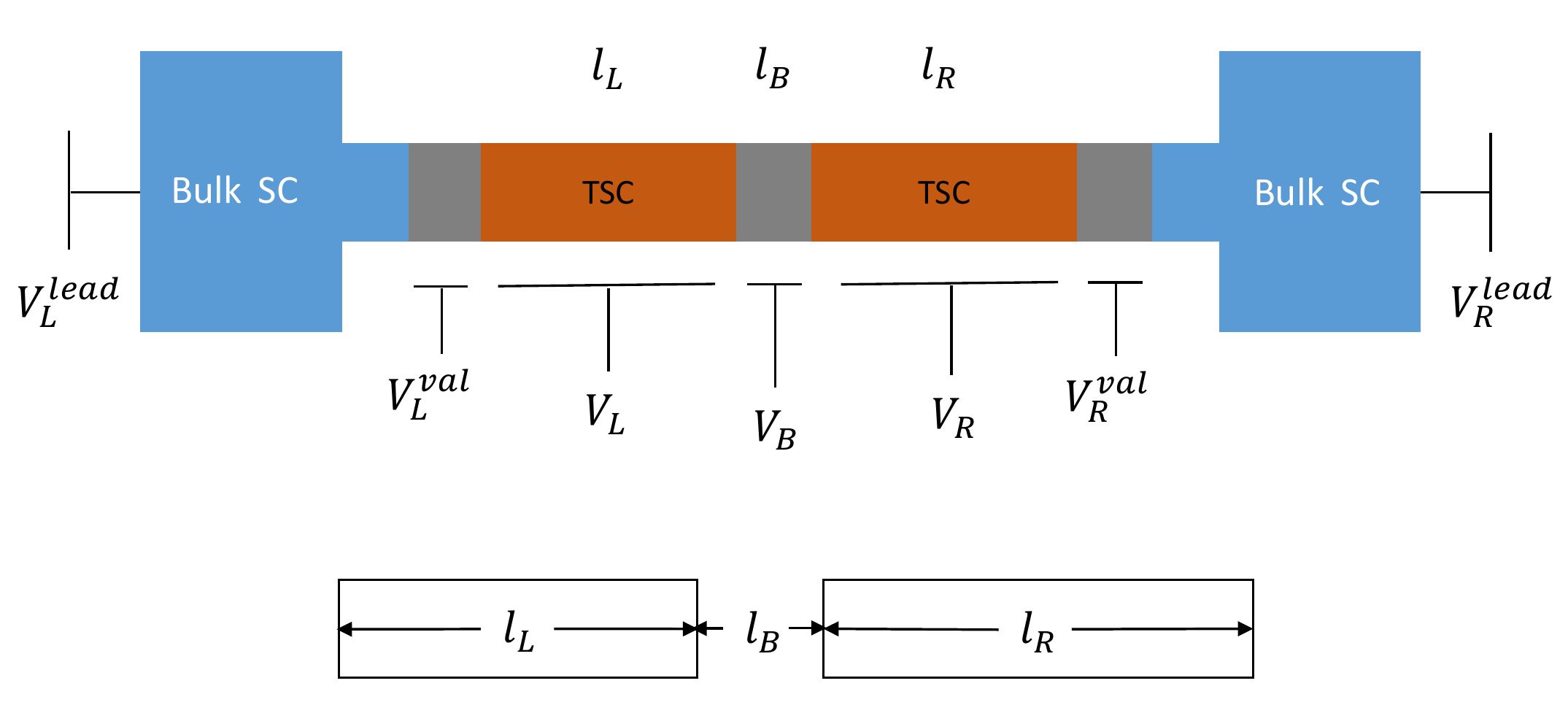}
\caption {Above, the schematic setup of the system that we study, where $V^{\rm lead}_{L,R}$ controls the bias between the leads, $V_{L,R}$ shifts the bands of the TSC, $V_B$ and $V^{val}_{L,R}$ tune the barrier and valves respectively. Below, a more basic version with the lengths of the two islands and tunnel barriers used in our calculations. Each TSC island is modeled as either a  Kitaev chain (Sec.~\ref{sec:Kitaev_chain}) or a 1D  SC semiconductor nanowire (Sec.~\ref{sec:semiconductor_model}).} \label{fig_schematic}
\end{figure}

\subsection{$\mu=0$}\label{subsec:mu=0}

\begin{figure}[!htbp]
\includegraphics[width=8.5cm]{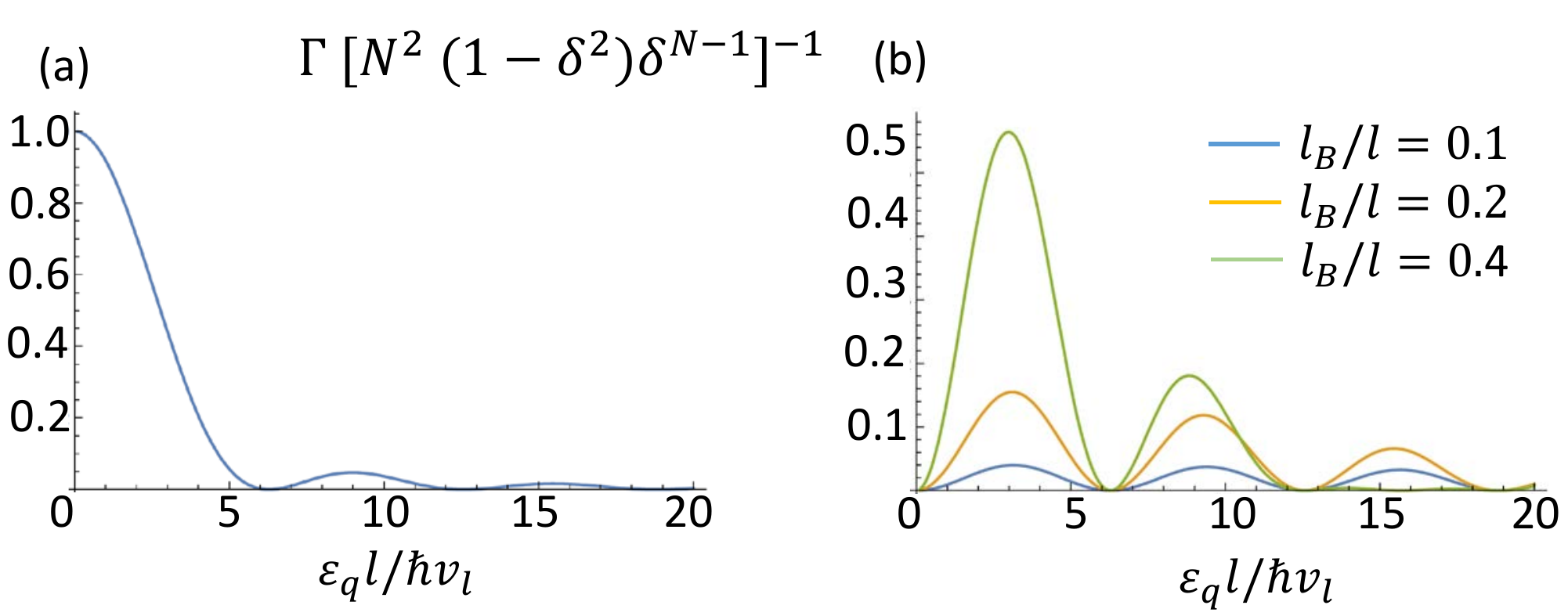}
\caption { The renormalized $\Gamma$ for $\mu=0$ [in Eq.~(\ref{eq:mu0})] as a function of phonon wavelength ($\sim 1/q_0=\hbar v_l/\varepsilon_q$) relative to the island lengths ($l$). Here $l$ is defined as, (a) the length of the shorter islands when $\delta^{|N_L-N_R|}\gg 1$, and (b) the length of each island when $\delta^{N_L-N_R}\approx 1$. $N=l/a_0$, the corresponding number of sites. $l_B$ is the length of the barrier.} \label{fig_mu0}
\end{figure}

At $\mu=0$, we have the following analytical solution for odd number of sites $N$  \cite{Schmidt_PRB12},
\begin{eqnarray}
b^\dag_0 \!=\! \frac{\sqrt{1\!-\!\delta^2}}{2}\! \sum^{(N-1)/2}_{i=0}\! \delta^i (c^\dag_{2i+1}\!+\! c_{2i+1} \!+\! c^\dag_{N-2i} \!-\! c_{N-2i}),
\end{eqnarray}
where $\delta=(\Delta-t)/(\Delta+t)$,  with $\varepsilon_0=0$, and the normalization factor $\sqrt{1\!-\!\delta^2}/2$ is precise at the large $N$ limit.  In this case,
\begin{eqnarray}
\Gamma &=& 4(1-\delta^2) \left|\delta^{(N_L-1)/2}\sum^{(N_L-1)/2}_{i=0} e^{ i q_0 R_{L,2i+1}}
 \right.
 \nonumber\\
&& \qquad \qquad
\left. -\delta^{(N_R-1)/2}\sum^{(N_R-1)/2}_{i=0} e^{ i q_0 R_{R,2i+1}}  \right|^2
\nonumber\\
&\approx& \frac{1-\delta^2}{a^2_0 q^2_0} \left|\delta^{(N_L-1)/2} (e^{iq_0 N_L a_0}-1)
 \right.
 \nonumber\\
&& \qquad \qquad
\left.
-\delta^{(N_R-1)/2}  e^{iq_0 l_B}(e^{iq_0 N_R a_0}-1)  \right|^2,
\label{eq:mu0}
\end{eqnarray}
where the summation is done by taking the long phonon wavelength continuum limit, $1/q_0\gg a_0$ (the distance between nearest sites). $\Gamma$ vanishes when the phonon wavelength ($2\pi/q_0$) is much less than the island length ($l_{L(R)}$), and is appreciable  when  $2\pi/q_0 \gtrsim l_{L(R)}$. $\Gamma$ depends most strongly (exponentially) on $N_L$ and $N_R$. If $\delta^{|N_L-N_R|}\gg 1$, then the right island terms are negligible. A renormalized $\Gamma$ is plotted in Fig.~(\ref{fig_mu0})(a) for this case.  If $\delta^{N_L-N_R}\approx 1$, $1/q_0 \lesssim l_B$ (the tunnel barrier length) is needed to ensure the left and right island terms are not mostly canceled out. In this situation, therefore $l_L\approx l_R\lesssim 1/q_0 \lesssim l_B$ is necessary for an appreciable coupling besides the $\delta^{N_{L(R)}}$ effect. For example, the $l_B/l=0.1$ curve in Fig.~\ref{fig_mu0} (d) shows that the phonon coupling is always largely suppressed for any $q_0$ if $l_{L(R)}\gg l_B$.

Note that the charge density of the Majorana mode is constant here,  $|u_{i,0}|^2 -|v_{i,0}|^2=4 \delta^{(N-1)/2}$ independent of site $i$, and is non-zero despite $\varepsilon_0$ being exactly at zero energy.

At $\mu=0$,  for even number of sites $N$  we have \cite{Miao_arxiv17},
\begin{eqnarray}
b^\dag_0 \!&\!=\!&\! e^{-v N}\sqrt{1\!-e^{-4v}} \sum^{N/2}_{i=1} (-1)^{(N+2)/2-i} \sinh [v(N\!+2\!-2i)]
 \nonumber\\
&& \times (c^\dag_{2i+1}\!+c_{2i+1}) \!-(\!-1)^{\!-\!N\!/2\!+i} \sinh (2vi) (c^\dag_{2i} \!-c_{2i}),\;\;
\end{eqnarray}
where $\frac{\sinh [v(N+2)]}{\sinh (vN)}=\frac{t+\Delta}{t-\Delta}$ (so $v\approx \frac{1}{2}ln\frac{t+\Delta}{t-\Delta}$), with $\varepsilon_0= (t+\Delta) (\frac{t-\Delta}{t+\Delta})^{N/2}$. Note that $|u_{i,0}|^2 -|v_{i,0}|^2=0$ even though the energy is not zero. Therefore in this case $\Gamma=0$, since the net electron density is always zero at every site.

\subsection{$t=\Delta$} \label{subset:t=Delta}

At $t=\Delta$, we have the following analytical solution (Ref.~\cite{Miao_arxiv17} with some changes),
\begin{eqnarray}
b^\dag_0\!& \!=\! &\! e^{-v N}\!\sqrt{1\!-e^{-2v}} \sum^{N}_{i=1} (-1)^{i+1} [\sinh [v(N\!+1\!-i)] (c^\dag_{i}\!+c_{i})
 \nonumber\\
&&
\qquad \qquad \qquad \qquad -\sinh (v i) (c^\dag_{i}-c_{i})],
\label{eq:b0_teqDelta}
\end{eqnarray}
where $\frac{\sinh [v(N+1)]}{\sinh (vN)}=\frac{2t}{\mu}$ (so $v\approx ln\frac{2t}{\mu}$), with energy $\varepsilon_0= 2t (\frac{\mu}{2t})^{N}$.  In this case,
\begin{eqnarray}
\Gamma \!\!\!&=\!\!& \!\!16(1\!-e^{\!-2v})\!^2  \! \left|e^{-2v N_L}\!\!\sum^{N_L}_{i=1}  e^{ i q_0 R_{L,i}} \!\sinh [v(N_L\!\!+1\!\!-i)]\!\sinh (v i)
\right.
 \nonumber\\
&& \left.\qquad \quad
- e^{-2v N_R}\!\!\sum^{N_R}_{i=1}  e^{ i q_0 R_{R,i}} \sinh [v(N_R\!+1\!-i)] \sinh (v i)   \right|^2
\nonumber\\
&\approx&  [1-(\frac{\mu}{2t})^2]^2 \left| F_L - {\rm sign}(\frac{-\mu}{t})^{N_L-N_R} e^{iq_0 l_B} F_R \right|^2,
\label{eq:Gamma_teqDelta}
\end{eqnarray}
where
\begin{eqnarray}
F\!\!_{X}\! =\! |\frac{\mu}{2t}|^{N\!\!_X} \![|\frac{\mu}{2t}|\! \frac{e^{iq_0 l_X}\!-\!1}{i q_0 a_0} \!+\! \frac{1}{i q_0 a_0\!+\!2\!\ln |\!\frac{\mu}{2t}\!|}\! -\! \frac{e^{iq_0 l_X}}{i q_0 a_0\!-\!2\!\ln |\frac{\mu}{2t}|}] ,\;
\end{eqnarray}
where we assume $\mu/2t$ is not very close to 1, desired for the topological phase, such that $(\mu/2t)^N{_{L(R)}}\ll 1$, to convert the sum to an integral. The sign oscillation of the Majorana wavefunction in Eq.~(\ref{eq:b0_teqDelta}) does not translate into $\Gamma$ except that for $\mu/t>0$ the relative sign  between the left and right island terms switches every time $N_L-N_R$ changes by one.

\begin{figure}[!htbp]
\includegraphics[width=8.5cm]{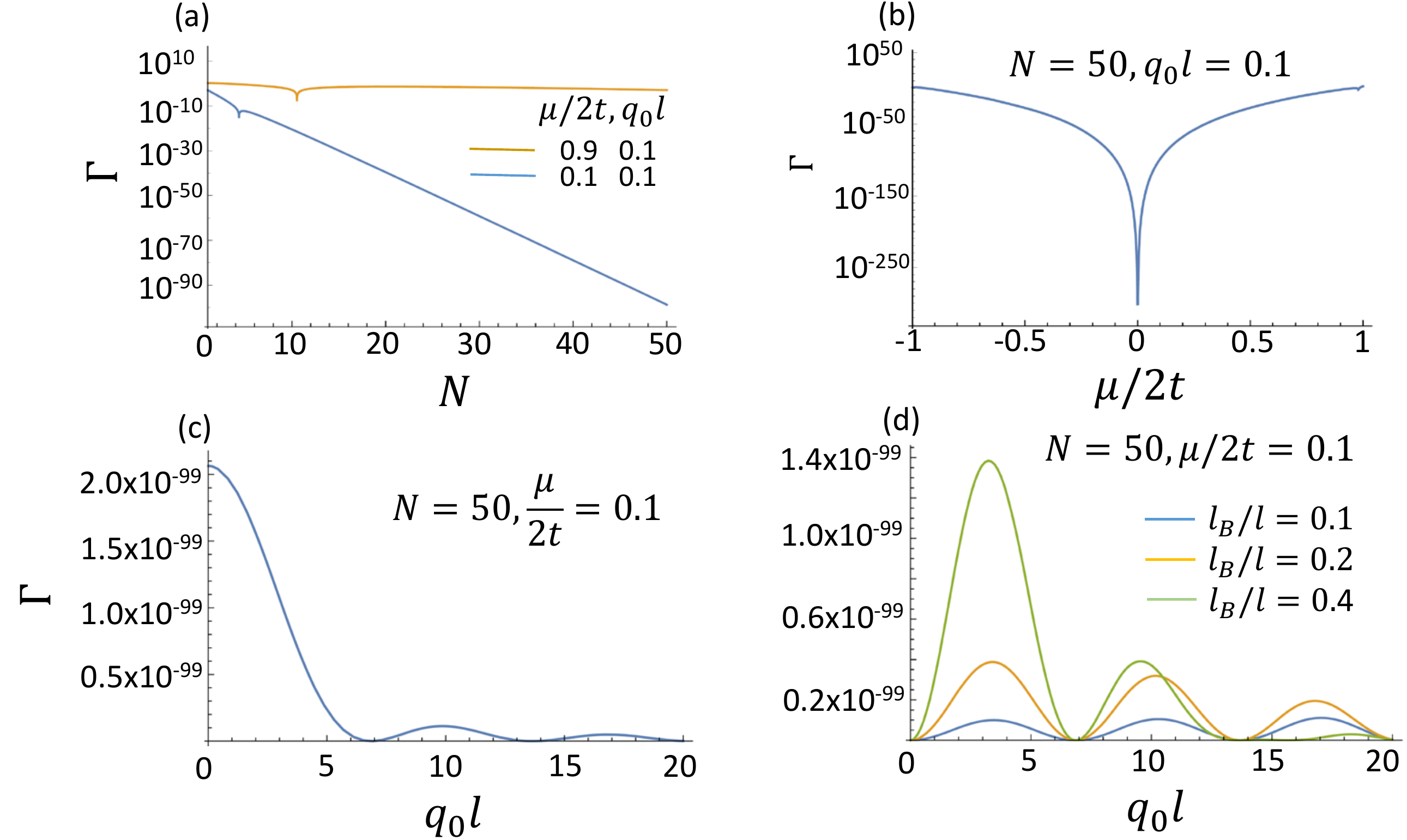}
\caption {$\Gamma$ for $t=\Delta$ [in Eq.~(\ref{eq:Gamma_teqDelta}) as a function of number of sites ($N$), the ratio ($\mu/2t$), and phonon wavelength relative to the island lengths ($q_0 l$). When $|\mu/2t|^{|N_L-N_R|}\ll 1$ only  the shorter island needs to be considered. In this case, we plot $\Gamma$ as a function of (a) island number of sites $N$, (b) $\mu/2t$, and (c) $q_0 l$ separately. (d) when $|\mu/2t|^{N_L-N_R}\approx 1$, we also show the effect of different barrier length $l_B$ on $\Gamma$ as the phonon wavevector varies.}\label{fig_Gamma_teqDelta}
\end{figure}

The representative results are plotted  in Fig.~\ref{fig_Gamma_teqDelta}. Again $\Gamma$ depends exponentially on $N_L$ and $N_R$. When the two islands are different in length, $|\mu/2t|^{|N_L-N_R|}\gg 1$, only the contribution from one island is important. In this case, we plot $\Gamma$ as a function of  number of sites ($N=l/a_0$), the ratio ($\mu/2t$), and phonon wavelength relative to the island lengths ($q_0 l$).
Figures~\ref{fig_Gamma_teqDelta} (a) and (b) show that the dominant behavior is determined by the $(\mu/2t)^N$ factor, except for the vanishing $\Gamma=0$ at some value of $N$ for $\mu/t>0$ cases. In these cases, the terms for each island in Eq.~(\ref{eq:Gamma_teqDelta}) can be factorized into a product including a real-number factor
\begin{eqnarray}
&&\frac{\mu}{2t} \frac{e^{iq_0 l}-1}{i q_0 a_0} + \frac{1}{i q_0 a_0+2\ln \frac{\mu}{2t}} - \frac{e^{iq_0 l}}{i q_0 a_0-2\ln \frac{\mu}{2t}}
\nonumber\\
&=& (e^{iq_0 l}+1) \left[\tan\frac{q_0 l}{2}\left(\frac{\mu/2t}{q_0 l/N}-\frac{q_0 l/N}{4\ln^2\frac{\mu}{2t}
 +(\frac{q_0 l}{N})^2} \right)
 \right. \quad
 \nonumber\\
&& \left.\qquad
+ \frac{2\ln \frac{\mu}{2t}}{4 \ln^2 \frac{\mu}{2t} +(q_0 l/N)^2}\right],
\end{eqnarray}
from which we can get the $N$ value (not necessarily an integer) that leads to $\Gamma=0$,
\begin{eqnarray}
N&=&- q_0 l \cot \frac{q_0 l}{2} \left[1+\sqrt{1+4\frac{\mu}{2t}(1-\frac{\mu}{2t})\tan^2\frac{q_0 l}{2}}\right]
 \nonumber\\
&& \qquad\left(4\frac{\mu}{2t}\ln\frac{\mu}{2t}\right)^{-1}.
\end{eqnarray}
Figure~\ref{fig_Gamma_teqDelta} (c) shows the suppression and oscillation of $\Gamma$ with $q_0 l$. Although for the case of $\mu/t>0$, the Majorana mode at $t=\Delta$ [Eq.~(\ref{eq:b0_teqDelta})] changes sign  for  every one or two lattice sites, no such rapid oscillation appears in the charge density, resulting in a smooth charge density whose coupling to the phonon mode is roughly $\propto (e^{iq_0 l}-1)/q_0 l$.
For $|\mu/2t|^{N_L-N_R}\approx 1$, the result is shown in Fig.~\ref{fig_Gamma_teqDelta}(d), analogous to Fig.~(\ref{fig_mu0})(b).

\subsection{general parameter choices}\label{subsec:Kitaev_general_para}

For parameter regimes other than the two subspaces constrained by $\mu=0$ and $t=\Delta$ respectively, one does not have any exact analytical solution. We present numerical results for general parameter choices in this section. Considering the eventual implementation in realistic semiconductor nanowires, we choose $-2t<\mu<0$ so as to simulate the conduction valley bottom scenario, and also take $\Delta\le t$ since the TSC pairing is usually much smaller than the hopping in any realistic system. There exist analytical approximations to the Majorana modes located at the end of semi-infinite wires \cite{Halperin_PRB12,DasSarma_PRB12}, which lead to qualitative features of the charge density profile $|u_{x,0}|^2 -|v_{x,0}|^2$ in a finite wire setup \cite{Lin_PRB12,BenShach_PRB15}. However, we do not find solutions that are quantitatively accurate enough over large parameter ranges.

\begin{figure}[!htbp]
\includegraphics[width=8.5cm]{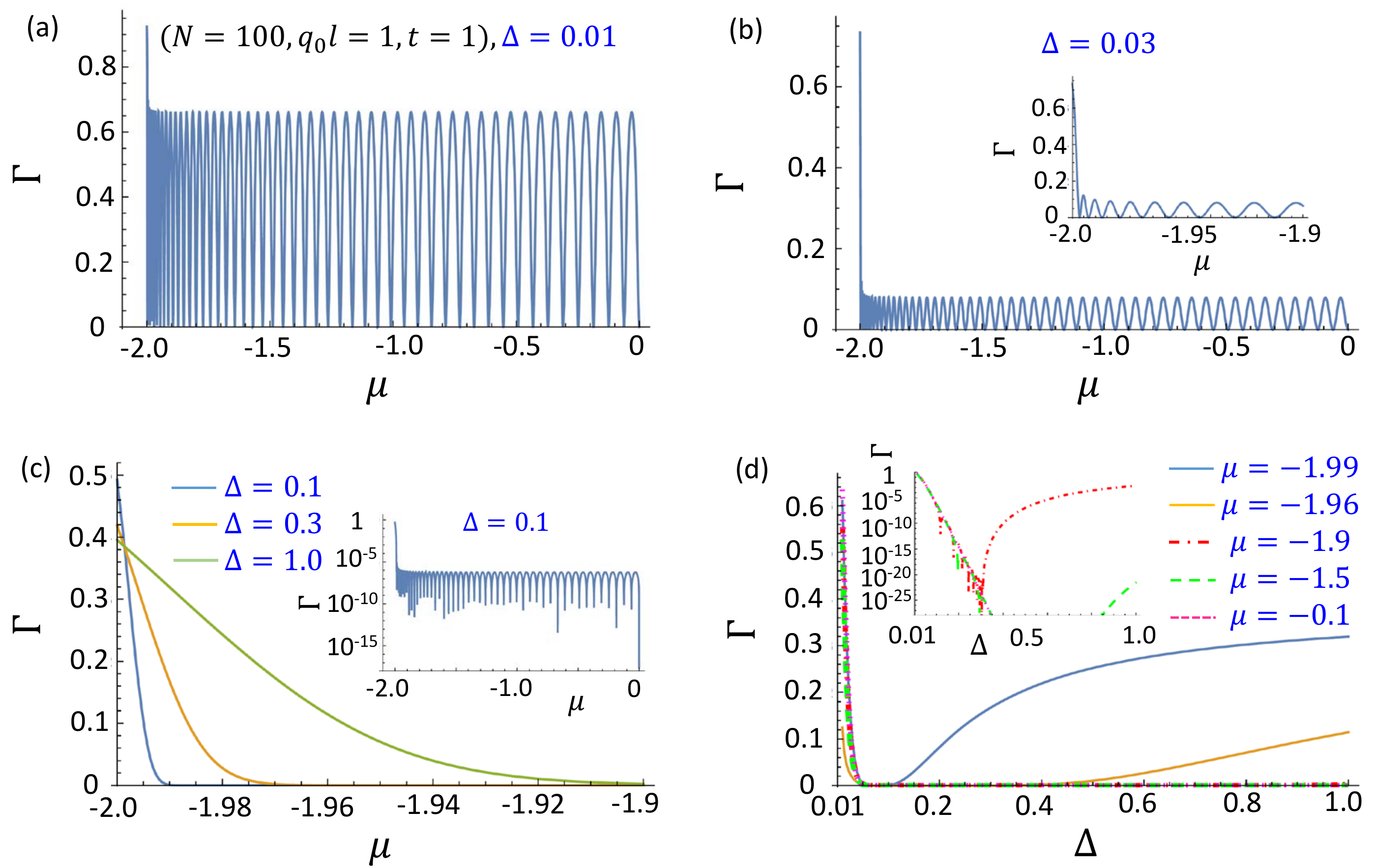}
\caption {$\Gamma$ for general $\mu$ and $\Delta$ values, with $t=1$, $N=100$ and $q_0 l=1$. Here we assume the contribution to Eq.~(\ref{eq:Gamma}) from the two islands is very different so only the dominant island is considered. (a)-(c) show $\Gamma$ as a function of $-2t<\mu\le0$ for several $\Delta=0.01$ (a), 0.03 (b), 0.1, 0.3 and 1 (c). The inset of (b) shows a close-up look of $\Gamma$ over $-2<\mu<-1.9$ for $\Delta=0.03$. (c) shows $\Gamma$ as $-2<\mu<-1.9$ for $\Delta=0.1,0.3,1$ where $\Gamma$ is visible. The inset of (c) shows the log plot of $\Gamma$ for $\Delta=0.1$. As the trend shown in (a), (b) and inset of (c), $\Gamma$ is even smaller for $\Delta=0.3$ and 1 when it is oscillatory over $\mu$ (not shown).
(d) shows $\Gamma$ as a function of $0<\Delta\le 1$ for several $\mu=-1.99,-1.96, -1.9, -1.5, -0.1$, and the inset shows the log plot for $\mu=-1.9,-1.5$ and $-0.1$ for which it is hard to visualize in the main plot of (d). For $\Gamma$ less than about $10^{-30}$, the numerical results becomes unreliable due to the large orders of magnitude difference between $|u_{0,i}|^2$, $|v_{0,i}|^2$ and $|u_{0,i}|^2-|v_{0,i}|^2$.
}\label{fig_Gamma_N100_q0l1}
\end{figure}

In Fig.~\ref{fig_Gamma_N100_q0l1}, we plot the calculated $\Gamma$ of Eq.~(\ref{eq:Gamma}) as a function of $-2<\mu\le0$ for several $\Delta$'s or as a function of $0<\Delta\le 1$ for several $\mu$'s, with  $t=1$, $N=100$ and $q_0 l=1$. The consequent numerical results for $\Gamma$ manifest two distinct trends. When $\mu+2t\rightarrow 0$ or $\Delta\rightarrow0$, the system approaches the topological phase boundary, where the MZM's energy deviation from zero grows, so does its effective finite charge content. As a result, we see that $\Gamma$, which is the integral over charge density enveloped by the phonon displacement, peaks toward unity at these two limits [in Figs.~\ref{fig_Gamma_N100_q0l1}(a)-(c) and (d), respectively].

Away from the topological boundary region, $\Gamma$  generally follows the variation of finite Majorana effective charge, for a relatively long phonon wavelength ($q_0 l=1$). The magnitude of the Majorana net occupation, net charge/$e=|\int (|u_{x,0}|^2 -|v_{x,0}|^2)dx|$, has been qualitatively estimated \cite{BenShach_PRB15} to be $\kappa l e^{-\kappa l}$, where $\kappa$ is the Majorana decay wavevector and $l$ is the effective MZM spatial separation. $\kappa$ approaches $(\mu+2t)/\Delta a_0$ (converted to our notations from Ref.~\cite{Pientka_NJP13})  when $\mu+2t\ll \Delta^2/2t$ (i.e., near the topological transition), and $\Delta /2t a_0$ when $\mu+2t\gg \Delta^2/2t$ (i.e., deep in the topological phase). This simple expression ($\kappa l e^{-\kappa l}$) captures the trends of $\Gamma$'s average amplitude, including the decreasing $\Gamma$ with $\mu$ in Figs.~\ref{fig_Gamma_N100_q0l1}(a)-(c) and the decrease followed by the rise of $\Gamma$ with $\Delta$ in Fig.~\ref{fig_Gamma_N100_q0l1}(d) (although Ref.~\cite{BenShach_PRB15} only studies the $\mu+2t\gg \Delta^2/2t$ regime, we find the expression applicable also into the $\mu+2t\ll \Delta^2/2t$ regime by directly comparing with our numerical work). $\kappa l e^{-\kappa l}$ peaks at $\kappa_0=1/l$, which is 0.01 for $a_0=1$ and $N=100$. Meanwhile for a given $\Delta$, as described in Ref.~\cite{Pientka_NJP13}, $\kappa$ increases (away from $\kappa_0$) with $\mu+2t$ and saturates near the boundary between  $\mu+2t\gg \Delta^2/2t$ and $\mu+2t\ll \Delta^2/2t$, to become $\kappa_M\sim\Delta/2ta_0$ which is $\Delta/2$ for $t=1$ and $a_0=1$. This dictates the fall of $\Gamma$ before its saturation in Figs.~\ref{fig_Gamma_N100_q0l1}(b) and (c). Too close to $\mu+2t=0$, we note however $\kappa$ is too small for Ref.~\cite{BenShach_PRB15} to be applicable. In  Fig.~\ref{fig_Gamma_N100_q0l1}(d) for a given $\mu$, the minimum of $\Gamma$ reflects the maximum of $\kappa$ around $\Delta^2/2t=\mu+2t$, $\kappa_M\sim\sqrt{(\mu+2t)/2t}/a_0>\kappa_0$. (In the Appendix, we also show the counterparts of Fig.~\ref{fig_Gamma_N100_q0l1} with $N=50$ or 200 in Fig.~\ref{fig_Gamma_N50-100-200_q0l1}, and  with $q_0l=3$ or 10 in Fig.~\ref{fig_Gamma_N100_q0l1-3-10}.)

However, the above expression misses the oscillation of $\Gamma$ as a function of $\mu$ deep in the topological phase [Figs.~\ref{fig_Gamma_N100_q0l1}(a),(b) and the inset of (c)]. This oscillation can be obtained by noting that the propagating wavevector $k$ does not need to satisfy the usual quantization rule $(n+\frac{1}{2})\frac{2\pi}{k}=l$ for the decaying Majorana modes $\gamma_{1,2}$, since the  boundary condition of vanishing wavefunction at one end is automatically satisfied (approximately, with an exponentially small correction $\sim e^{-\kappa l}$). As a result, $\gamma_{1,2}$ in a finite wire may follow the analytical solutions in a semi-infinite wire. Following that, a simple connection to Eq.~(18) of Ref.~\cite{BenShach_PRB15} can be made by using $\sin(k_F x) \sin(k_F(l-x)) = \frac{1}{2} [\cos (k_F(2x-l))-\cos (k_F l)]$, where $\cos (k_F l)$ oscillates with $k_F$ rather than being fixed to 1 or $-1$. That is, not only the charge density distribution oscillates in real space, the mean value of the oscillation also oscillates as a function of $k_F$ (the smooth oscillation of net charge with $\mu$).  We have, for $k_F>\kappa > 1/l$,
\begin{eqnarray}
|\!\int |u_{x,0}|^2 \!-\!|v_{x,0}|^2dx \!|
\!&\!\approx\!&\! 2\kappa e^{\!-\!\kappa l}\! |l \cos(k_F l)\! -\! \frac{1}{k_F} \sin(k_F l)|
\nonumber\\
&\approx& 2\kappa e^{-\kappa l} l |\cos(k_F l)| .
\label{eq:charge}
\end{eqnarray}
On the other hand, when  $\kappa l\ll 1$, the net charge approaches discrete sign switch at $k_F =n \pi/l$.

As a result, it is \textit{interesting} to tune $\mu$ to the spots of vanishing $\Gamma$, to suppress Majorana qubit relaxation (and decoherence) induced by phonon, as well as by charge noise. That is, the relaxation is very sensitive and periodic in terms of $\mu$. The fact that the MZM splitting oscillations due to wavefunction overlap leads to specific $\mu$ values (or equivalently, separation $l$ values) where the system is topologically protected (i.e. $\Gamma=0$) in spite of the inter-MZM separation not being very large is of course already implicit in the early works pointing out MZM oscillations \cite{Cheng_PRL09, Cheng_PRB10, DasSarma_PRB12}.   In the current work, the direct effect of these oscillations \cite{Cheng_PRL09, Cheng_PRB10, DasSarma_PRB12} on phonon relaxation is showing up.

Note that the oscillation of net charge versus $k_F$, in most relevant cases, has a phase difference of $\pi$ from the oscillation of the Majorana energy splitting. The latter has been studied in Ref.~\cite{DasSarma_PRB12}. In fact, using the derivation from the appendix of Ref.~\cite{DasSarma_PRB12}, we find that the dominant energy term in the regime where $k_F\gg \kappa$ (the condition for oscillation) is proportional to $\sin(k_F l)$, rather than the $\cos(k_F l)$ concluded in Ref.~\cite{DasSarma_PRB12} [and an additional factor of 2; we have numerically checked the Majorana energy solved from Eq.~(\ref{eq:H_BdG_nanowire}) in Sec.~\ref{sec:semiconductor_model}]: Using our Kitaev parameters, and Eqs.~(25) and (29) in Ref.~\cite{DasSarma_PRB12}, by plugging in $u_l\propto e^{-\kappa x} \sin(k_F x)$, we have
\begin{eqnarray}
|\varepsilon_o-\varepsilon_e|\!&\!\propto \!&\! a_0 t e^{-\kappa l} \sin( \frac{k_Fl}{2}) [-\kappa \sin( \frac{k_Fl}{2}) + k_F \cos( \frac{k_Fl}{2}) ]
\nonumber\\
&&+ \Delta e^{-\kappa l}  \sin^2( \frac{k_Fl}{2})
\nonumber\\
&\approx& \frac{a_0 t k_F}{2} e^{-\kappa l} \sin(k_F l).
\end{eqnarray}
This is essentially due to the derivative operator in  (29) of Ref.~\cite{DasSarma_PRB12}, and the $\mu+2t\ll \Delta^2/2t$ condition needed in the oscillation regime.
So the net charge reaches antinodes (local maximum) when the Majorana modes reach nodes in the energy (the degeneracy spots) as functions of $k_F$ or $\mu$, and vice versa. We have numerically verified these conclusions for the Kitaev model.

\begin{figure}[!htbp]
\centering
  \begin{tabular}{c}
  \includegraphics[width=8.5cm]{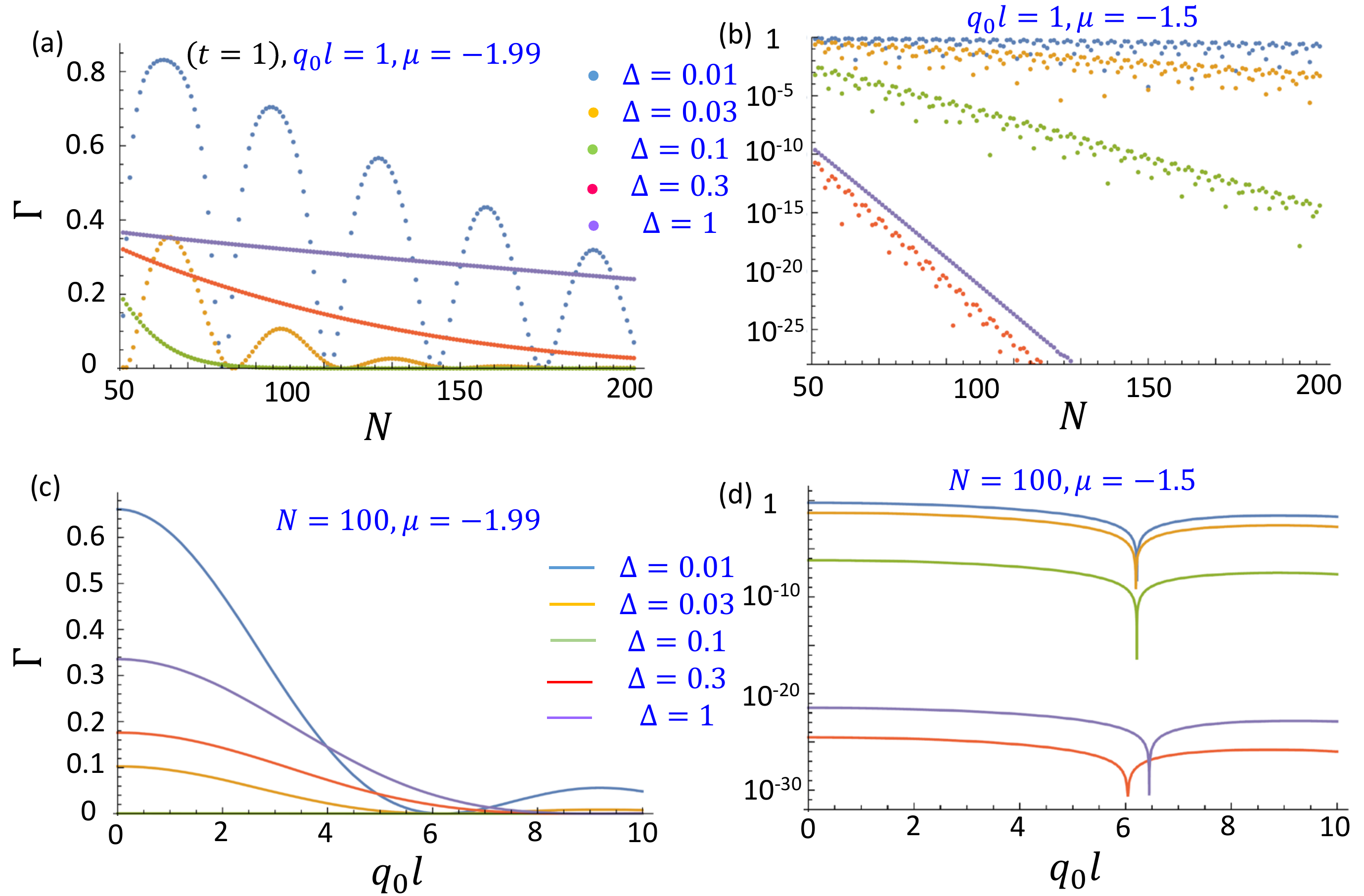} \\
  \includegraphics[width=8.5cm]{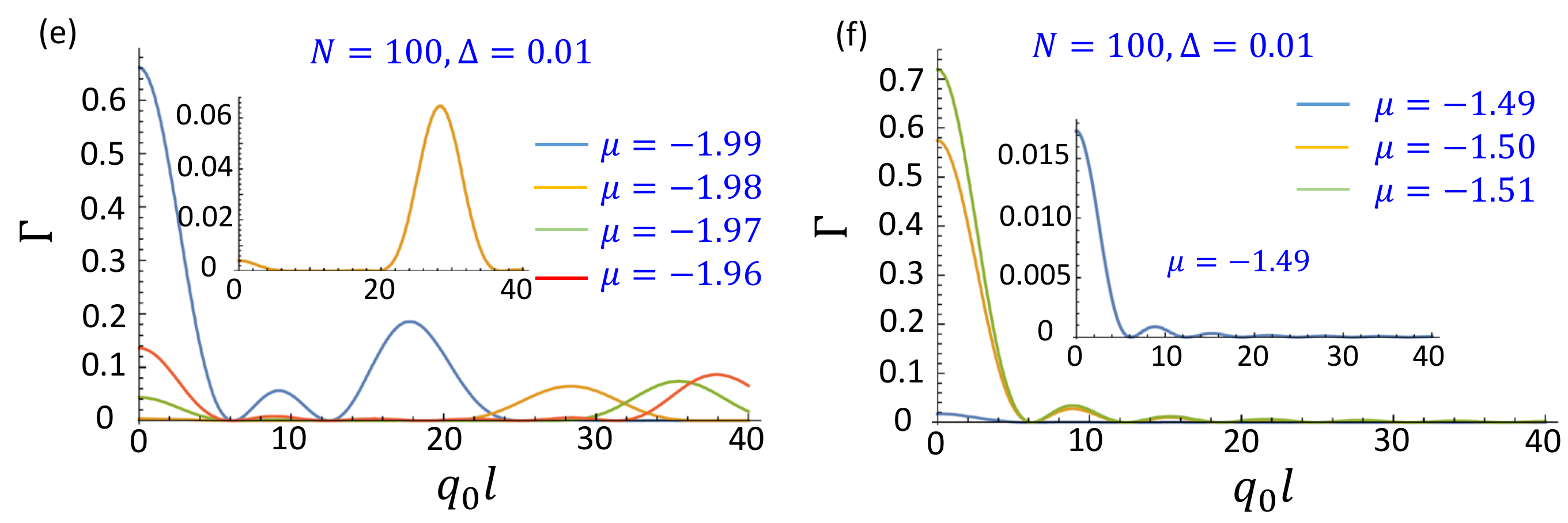}
  \end{tabular}
\caption {$\Gamma$ as a function of $N$, or as a function of  $q_0 l$, for representative  $\mu$ and $\Delta$. $t=1$. One of the two islands is assumed to dominate $\Gamma$ in Eq.~(\ref{eq:Gamma}). $\Gamma(N)$ as $50<N<200$ for (a) $\mu=-1.99$ or (b) $\mu=-1.5$ at $q_0l=1$, and $\Gamma(q_0 l)$ as $0<q_0l<10$ for (c) $\mu=-1.99$ or (d) $\mu=-1.5$ at $N=100$, and at $\Delta=0.01$, 0.03, 0.1, 0.3 or 1. $\Gamma(q_0 l)$ as $0<q_0 l<40$ for (e) $\mu=-1.99, -1.98, -1.97$ or $-1.96$, or (f) $\mu=-1.49$, $-1.5$, or $-1.51$ at $N=100$ and  $\Delta=0.01$. A more detailed view of $\mu=-1.98$ and $\mu=-1.49$ is also shown in the inset of (e) and (f) respectively.
}\label{fig_Gamma_mu199-150_Delta001-1}
\end{figure}

Under the same level of approximation, we derive analytical expressions for $\Gamma$ [Eq.~(\ref{eq:Gamma})]. Using $|u_{x,0}|^2 -|v_{x,0}|^2 \approx -4\kappa  e^{-\kappa l} \sin(k_F x) \sin(k_F(l-x))$,
\begin{eqnarray}
\Gamma&\approx & 4\left|G_L-
 e^{iq_0 l_B}G_R \right|^2,
\end{eqnarray}
where
\begin{eqnarray}
G_{i}\! =\!\kappa_i e^{-\kappa_i l_i} \!\!\int_0^{l_i}\! dx [\cos(k_{F,i} (2x-l_i))\!- \!\cos(k_{F,i} l_i)] e^{iq_0 x}.\;
\end{eqnarray}
 When one of the islands dominates,
\begin{eqnarray}
\Gamma\!\!&\!\approx\!&\!\! 4\kappa^2 e^{\!-2\kappa l}\! \left|\!\frac{i(e^{iq_0l}\!-\!1) q_0 \!\cos(k_F l)\!+\! 2(e^{i q_0 l}\!+\!1)  k_F \!\sin(k_F l)}{(2k_F-q_0) (2k_F+q_0)}
 \right.
 \nonumber\\
&& \left.\qquad\qquad
- \cos(k_F l)\frac{e^{i q_0 l}-1}{i q_0}\right|^2
\nonumber\\
&\approx& \left[4 \frac{\kappa}{q_0} e^{-\kappa l} \cos(k_F l) \sin(\frac{q_0 l}{2})\right]^2 , \label{eq:Gamma_1island}
\end{eqnarray}
where the last approximation holds for $k_F\gg \{q_0, 1/l\}$. When the two islands contribute the same except for the phase difference with the phonon $(e^{iq_0 l_B})$, one has
\begin{eqnarray}
\Gamma\approx \left[8 \frac{\kappa}{q_0} e^{-\kappa l} \cos(k_F l) \sin(\frac{q_0 l}{2}) \sin(\frac{q_0 l_B}{2})\right]^2.\label{eq:Gamma_equal_islands}
\end{eqnarray}

Next we relax constraints on other independent degrees of freedom that $\Gamma$ depends on. They are the number of sites ($N$) and the phonon wavevector relative to the island length ($q_0 l$). Then without loss of generality, the hopping $t$ can be fixed to unity. Still we assume one of the two islands is dominant due to the strong exponential dependence on the island length. Figures~\ref{fig_Gamma_mu199-150_Delta001-1} (a)-(d) show $\Gamma$ as a function of $50<N<200$ or as a function of $0<q_0 l<10$  respectively, for several representative $\mu$ and $\Delta$. (In the Appendix, one can find in Figs.~\ref{fig_Gamma_N50-100-200_q0l1}  and \ref{fig_Gamma_N100_q0l1-3-10} the complementary results to Fig.~\ref{fig_Gamma_mu199-150_Delta001-1}.)

Figure~\ref{fig_Gamma_mu199-150_Delta001-1} (a) shows the calculated $\Gamma(N)$ for $\Delta=0.01, 0.03, 0.1, 0.3$ and 1 for fixed $\mu=-1.99$ and $q_0 l=1$. From the decay wavevector trends provided previously, for a given $\mu$ the fastest decay happens around $\Delta^2/2t\sim \mu+2t$, which is close to the $\Delta= 0.1$ curve. For smaller $\Delta$, oscillation emerges obeying Eq.~(\ref{eq:charge}) with wavevector $k_F=\sqrt{(\mu+2t)/t}$. Figure~\ref{fig_Gamma_mu199-150_Delta001-1} (b), with $\mu=-1.5$ instead, puts all five curves into the fast decay regime, $\kappa=\Delta/2$ with each given $\Delta$ (though not quite so for $\Delta=1$, which is near the boundary of $\Delta^2/2t\approx \mu+2t$).

Figures~\ref{fig_Gamma_mu199-150_Delta001-1} (c) and (d) show the typical dependence on $q_0 l$ when the chemical potential is near the band bottom ($\mu=-1.99$) or deep into the band ($\mu=-1.5$). As we discussed before, when the charge density profile $|u_0(x)|^2-|v_0(x)|^2$ is mostly positive or negative, which is the case for $\mu=-1.99$ or for $\mu=-1.5$ at $N=100$, the dependence of $\Gamma$ on $q_0$ follows roughly $\propto (e^{iq_0 l}-1)/q_0 l$. For these cases, the barrier effect for two nearly identical islands is analogous to that in Figs.~\ref{fig_mu0}(b) and \ref{fig_Gamma_teqDelta}(d), and therefore is not shown.

The dependence of phonon wavevector $q_0$ is more important and dramatic for the regime of vanishing net charge. These vanishing charge spots lead to the periodic suppression of $\Gamma$ shown in Figs.~\ref{fig_Gamma_N100_q0l1} (a)-(c) and also in Fig.~\ref{fig_Gamma_mu199-150_Delta001-1}(a). As discussed before, at these special parameter values, the charge density profile oscillates around zero, and a close proximity of the charge density wavevector and the phonon wavevector would greatly enhance the coupling. We show this effect in Fig.~\ref{fig_Gamma_mu199-150_Delta001-1}(e)  by tuning $\mu$ slightly around $-1.99$. Note the curve for $\mu=-1.98$, which is tuned to have nearly vanishing $\Gamma$ (and so is the relaxation) at long phonon wavelength limit $q\rightarrow0$, is enhanced by an order of magnitude when $q_0\sim l/30$ matches $2k_F$. In Fig.~\ref{fig_Gamma_mu199-150_Delta001-1} (f), by tuning $\mu$ around $-1.5$ we find $\Gamma$ is suppressed at about $\mu=-1.49$ for $q\rightarrow0$. However, due to the very large $k_F$ which no phonon wavevector matches in the shown range of $0<q_0l<40$, the dependence again follows roughly $\propto (e^{iq_0 l}-1)/q_0 l$. It is possible to have a large $q_0$ (or $\varepsilon_{q_0}$) by applying a large potential bias between the two islands.

\begin{figure}[!htbp]
  \includegraphics[width=8.5cm]{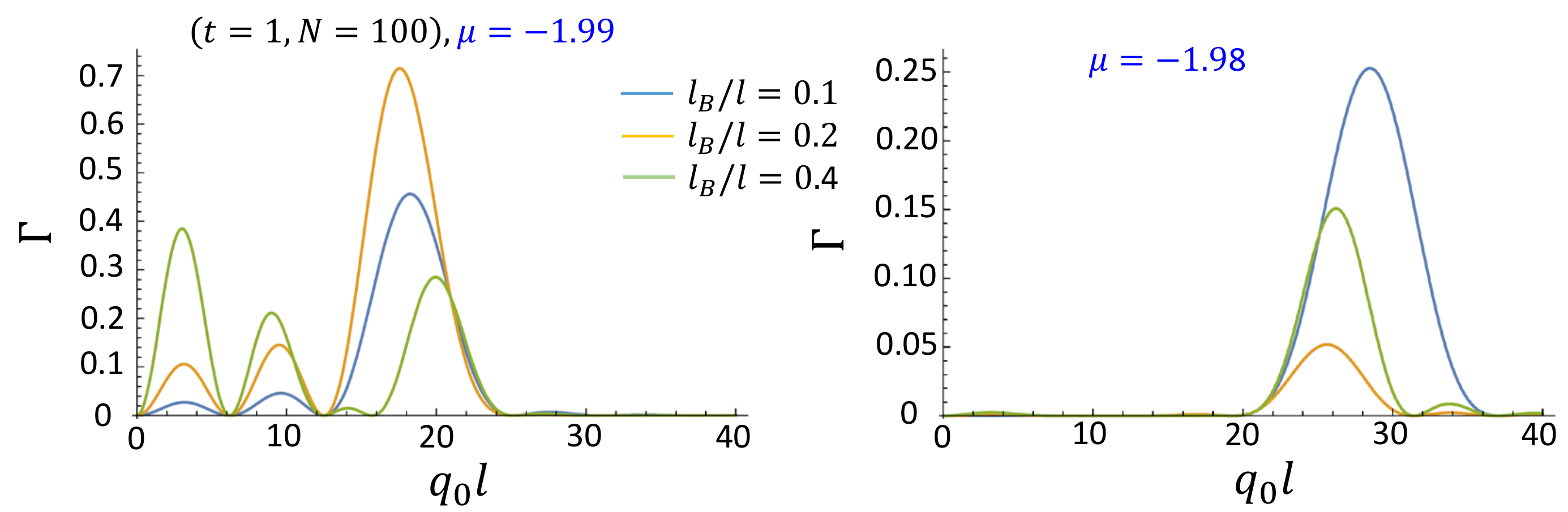}
\caption {$\Gamma$ as a function  $q_0 l$, for  (a) $\mu=-1.99$ or (b) $\mu=-1.98$ with different barrier length $l_B=0.1 l, 0.2 l$ and $0.4 l$, when the two islands are identical. $t=1, N=100$ and $\Delta = 0.01$.
}\label{fig_Gamma_N100_mu198-199_barrier01-02-04}
\end{figure}

The case of vanishing net charge also makes the effect of tunnel barrier special, particularly when the two islands are nearly identical. In Figs.~\ref{fig_mu0}(b) and \ref{fig_Gamma_teqDelta}(d) or other cases with the charge density mostly of the same sign, and for the experimentally relevant  $l_B/l\ll 1$, $\Gamma$ is always much less than that from each island in the strongest relaxation regime (that is, the long phonon wavelength regime). This is because $|1-e^{i q_0 l_B}|\ll 1$ in this regime. However, for the case in the inset of Fig.~\ref{fig_Gamma_mu199-150_Delta001-1}(e), $\mu=-1.98$, the strongest phonon coupling regime (the resonance) is not suppressed by the barrier for $l_B/l\ll 1$, since $q_0 l_B$ can be the order of one or greater. We show  in Fig.~\ref{fig_Gamma_N100_mu198-199_barrier01-02-04} the variation of $\Gamma(q_0 l)$ for different $l_B/l$, contrasting between $\mu=-1.99$ and $\mu=-1.98$ cases. In comparison with Fig.~\ref{fig_Gamma_mu199-150_Delta001-1} (e), it is clear that the strongest peak for  the $\mu=-1.98$ curve largely remains at $l_B/l=0.1$ in Fig.~\ref{fig_Gamma_N100_mu198-199_barrier01-02-04} (b), and is largely reduced for the $\mu=-1.99$ curve  at $l_B/l=0.1$ in Fig.~\ref{fig_Gamma_N100_mu198-199_barrier01-02-04} (a) though the smaller peak at $q_0 l\approx18$ is preserved.

\section{Superconducting spin-orbit nanowire model}\label{sec:semiconductor_model}

We choose  materials parameters associated with InAs or InSb nanowires used in the experiments (see Table I) \cite{Mourik_Science12, Rokhinson_NatPhys12, Deng_NanoLett12, Churchill_PRB13, Das_NatPhys12, Finck_PRL13, Albrecht_Nature16, Deng_Science17, Zhang_Nature18}. Now the summation over $i$ in Eq.~(\ref{eq:Gamma}) includes spins.  The one-dimensional BdG Hamiltonian is as usual,
\begin{eqnarray}
H_{\rm BdG}\!=\! (-\!\frac{\hbar^2}{2m^*}\frac{\partial^2}{\partial x^2}\!-\!\mu)\tau_z \!+\! V_z \sigma_z \tau_z \!+\! i \alpha \frac{\partial}{\partial x}\sigma_y \tau_z \!+\! \Delta \sigma_y \tau_y,\;\; \label{eq:H_BdG_nanowire}
\end{eqnarray}
where $\alpha$ is the nanowire spin-orbit coupling, $V_z$ is the Zeeman spin splitting, $\sigma$ and $\tau$ are the Pauli matrices in spin and particle-hole space, respectively. Note that the reference zero for $\mu$ is redefined as the band bottom in the absence of $E_z, \alpha$ and $\Delta$, and $\Delta$ is the proximity-induced $s$-wave SC pairing rather than the $p$ pairing in the Kitaev model of Sec.~\ref{sec:Kitaev_chain}.

\begin{table} [!htbp]
\caption{ Materials parameters for the BdG Hamiltonian in Eq.~(\ref{eq:H_BdG_nanowire})\cite{Lutchyn_arxiv17}.
}
\tabcolsep=0.2 cm
\begin{tabular}{ccccc}
 \hline\hline
            &  $m^*$  &  $\alpha$ & $\Delta$  & $V_z$ \\
\hline
InAs    &   0.023 $m_e$       &  0.2-0.8 eV\AA  & \multirow{2}{*}{0.1-1 meV} &\multirow{2}{*}{0-10 meV} \\
InSb    &   0.014 $m_e$       &  0.2-1 eV\AA   \\
\hline\hline
\end{tabular}
\end{table}

In comparison with the Kitaev model studied above, instead of three parameters $t, \mu, \Delta$ (two independent ones), we have now five parameters, $m^*, \alpha, \Delta, V_z, \mu$, in addition to the recurring dependence on $l, q_0 l$ and $q_0 l_B$ involving phonon and two-island geometry. Next we will probe the dependence of the dimensionless Majorana-phonon coupling $\Gamma$ on this new set of parameters.

Compared with the Kitaev chain model, since the $\mu=0$ limit (Sec.~\ref{subsec:mu=0}) in the Kitaev model corresponds to the mid-band chemical potential that does not happen in the semiconductor, we do not have a similar limit to study. Similarly the $t=\Delta$ limit in Sec.~\ref{subset:t=Delta} does not apply in the semiconductor nanowire.

Now we study the independent degrees of freedom in the space of five parameters $m^*, \alpha, \Delta, V_z, \mu$ in regard to determining $\Gamma$. First, it is easy to see, as in the Kitaev case, if we change all $1/m^*, \alpha, \Delta, V_z, \mu$ proportionally, the eigenvector states of the tight-binding matrix do not change, and consequently neither does $\Gamma$ (for  a fixed set of $l, q_0, l_B$). Therefore, we may just study the InSb case, which has a smaller $m^*=0.014 m_e$ and thus covers a greater range of the effective parameters than studying the InAs case. Also, the precise materials parameters for the realistic superconductor-semiconductor hybrid nanowire system are unknown anyway (here we are using the known band structure parameters for the nanowire in isolation) since the superconductor most likely strongly renormalizes the nanowire band parameters including $m^*, \alpha, \Delta$, and the Land\'{e} $g$-factor (and hence $V_z$) as is well-established theoretically \cite{Cole_PRB15, Cole_PRB16, Stanescu_PRB17, Woods_arxiv18}.

The spatial distribution of the Majorana wavefunction is strictly determined by four degrees of freedom. However, the main physical dependence can be captured by two quantities, the decaying and oscillating wavelengths, or wavevectors $\kappa$ and $k_F$, just as in the Kitaev case (while the Kitaev model only has two independent parameters to start with, here we have four independent parameters). This applies to a not too small $V_z$, that is, $V_z$ is not too close to the critical topological quantum phase transition value $V_{z,c}$. On the contrary when $V_z$ is very close to $V_{z,c}$, a very small $w$ dominates the decay wavevector \cite{Klinovaja_PRB12, DasSarma_PRB12}. In general, there are these three wavevectors at play, in addition to a fourth constant coefficient of order unity involved in the charge distribution \cite{DasSarma_PRB12}. One  is  usually interested in $V_z$  far above $V_{z,c}$, and the charge distribution is approximately described by Eq.~(\ref{eq:charge}) again in this strong-field topological limit.

When either $\alpha$ or $\Delta$ is smaller (so is the induced BdG gap) than $V_z$, one has \cite{DasSarma_PRB12}
\begin{eqnarray}
k_F \!\!&\!\approx\!&\! \frac{\!\sqrt{2m^*}}{\hbar}\!\sqrt{\!\!\sqrt{(\mu\!+\!\frac{ m^*\alpha^2}{2\hbar^2})^2 \!+\!(V_z^2 \!-\!\Delta^2 \!-\!\mu^2)} \!+\! (\mu\!+\!\frac{ m^*\alpha^2}{2\hbar^2})}, \label{eq:kF_semi_wire_1}
\nonumber\\
\\
\kappa\!&\!\approx\!&\! \Delta \alpha m^* \bigg/\hbar^2 \sqrt{(\mu+\frac{ m^*\alpha^2}{2\hbar^2})^2 +(V_z^2 -\Delta^2 -\mu^2)}.
\label{eq:kappa_semi_wire_1}
\end{eqnarray}
This can be further approximated, away from the topological transition boundary ($V_z/V_{z,c}\gg 1$) and at $V_z \gg E_{\rm so}$ where $E_{\rm so}=\frac{ m^*\alpha^2}{2\hbar^2}$,
\begin{eqnarray}
k_F &\approx&  \frac{\sqrt{2m^*V_z}}{\hbar}, \label{eq:kF_semi_wire_2}
\\
\kappa&\approx& \frac{\Delta \alpha m^*}{\hbar^2   V_z}.\label{eq:kappa_semi_wire_2}
\end{eqnarray}
We have checked the numerical wavefunctions and verified the above approximate dependence of the decay and oscillatory wavevectors.

Several relevant features stem from Eqs.~(\ref{eq:kappa_semi_wire_1})-(\ref{eq:kappa_semi_wire_2}): For $V_z\gg \{\mu, \Delta, E_{\rm so}\}$, $k_F$ only depends on $V_z$ ($k_F\propto \sqrt{V_z}$) and not on $\Delta, \alpha$ or $\mu$, while $\kappa$ is proportional to $\Delta, \alpha$ and $1/V_z$. So there are two independent parameters, $V_z$ and $\Delta\alpha$ ($m^*$ is assumed fixed here). When $\mu$ is not small, it  affects mostly $k_F$, roughly as $\propto\sqrt{\mu}$, but not $\kappa$. When $\Delta$ is not small, it affects $k_F$ by adding $-\Delta^2$ to $V_z^2$. Finally, unless $E_{\rm so}$ is comparable to both $V_z^2 -\Delta^2 -\mu^2$ and $|\mu|$, which is unlikely, no additional effect comes from $\alpha$.

\begin{figure}[!htbp]
\centering
  \begin{tabular}{c}
  \includegraphics[width=8.5cm]{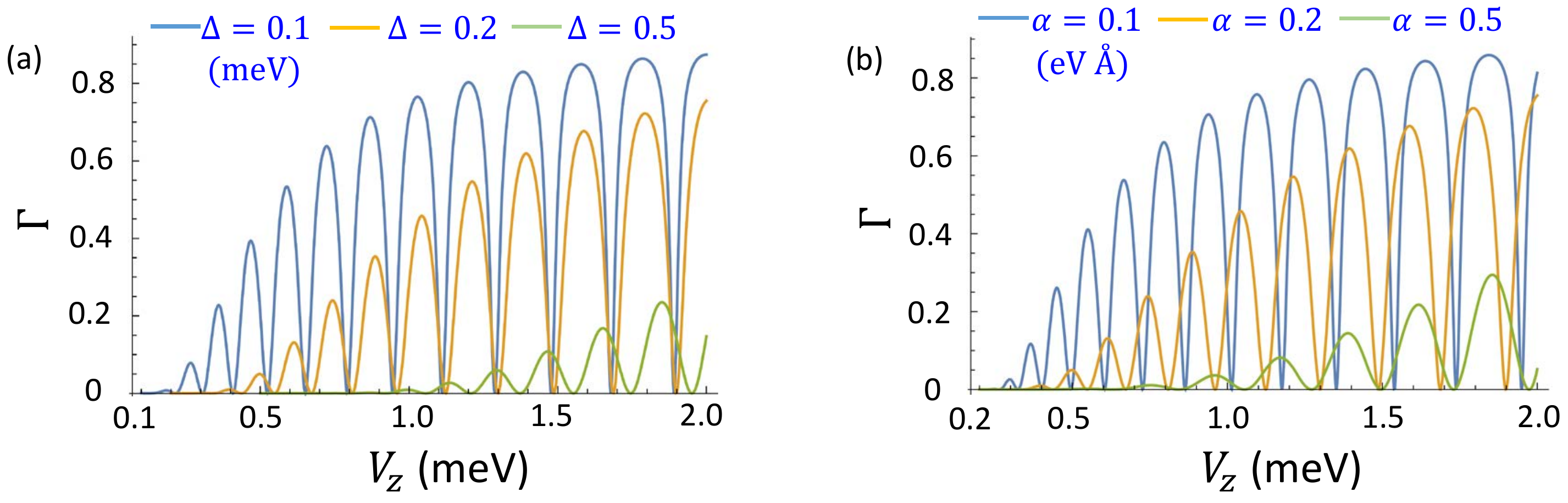} \\
  \includegraphics[width=8.5cm]{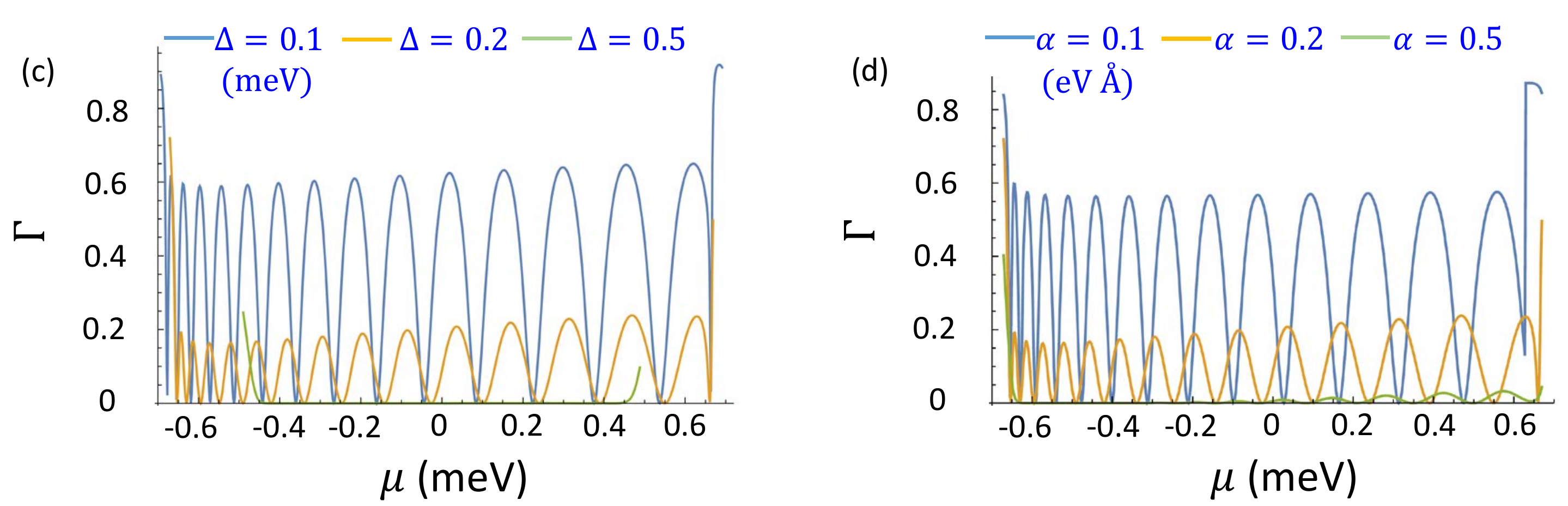} \\
  \includegraphics[width=8.5cm]{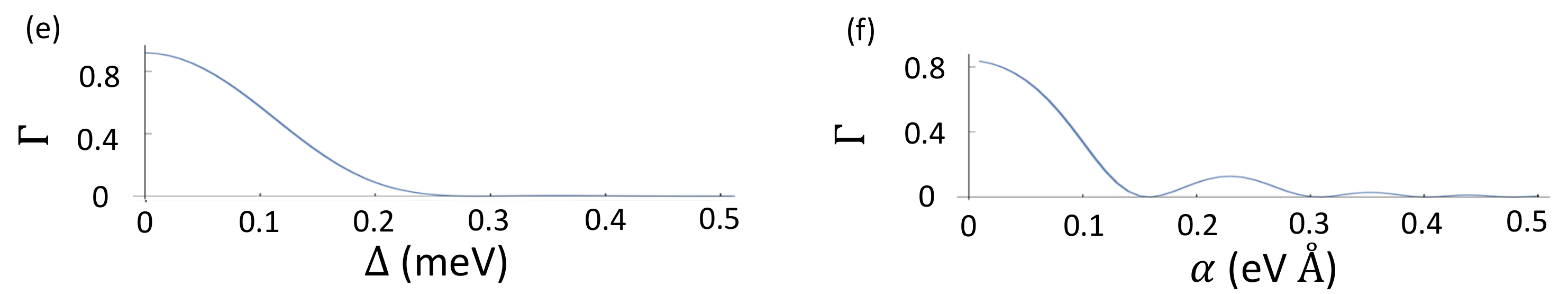}
  \end{tabular}
\caption { The core Majorana-phonon interaction function $\Gamma$ in a two-island InSb nanowire with proximity-induced superconductivity. Here one of the islands is assumed to dominate $\Gamma$, with $m^*=0.014 m_e$ and $l=2 ~\mu$m, and we show results with a fixed $q_0=1/l$. (a) $\Gamma(V_z)$ with fixed $\mu=0, \alpha=0.2$ eV\AA \,and three different $\Delta=0.1,0.2$ and $0.5$ meV. (b) $\Gamma(V_z)$ with fixed $\mu=0, \Delta= 0.2$ meV  and three different $\alpha=0.1,0.2$ and $0.5$ eV \AA. (c) $\Gamma(\mu)$ with fixed $V_z=0.7$ meV, $\alpha=0.2$ eV\AA\, and three different $\Delta=0.1,0.2$ and $0.5$ meV. Note that most of the $\Delta=0.5$ curve is too low to be visible. (d) $\Gamma(\mu)$ with fixed $V_z=0.7$ meV, $\Delta= 0.2$ meV  and three different $\alpha=0.1,0.2$ and $0.5$ eV \AA. (e) $\Gamma(\Delta)$ with $\alpha=0.2$  eV\AA\ and (f) $\Gamma(\alpha)$ with $\Delta=0.2$ meV,  with fixed $\mu=0$ and $V_z=0.7$ meV.
}\label{fig_Gamma_semi_fixed1island}
\end{figure}

Now we can study the quantity $\Gamma$, and compare the results with the analytical expressions in Eqs.~(\ref{eq:Gamma_1island}) and (\ref{eq:Gamma_equal_islands}).
We plot representative dependence of $\Gamma$ on various parameters. Many of the discussions in Sec.~\ref{subsec:Kitaev_general_para} also apply here. We will not repeat them while referring to some parts when necessary.  In Fig.~\ref{fig_Gamma_semi_fixed1island} we consider one island dominating the result, and consider a fixed length $l=5 ~\mu$m and phonon wavevector $q_0=1/l$. First, in Figs.~\ref{fig_Gamma_semi_fixed1island} (a) and (b) we focus on the dependence on $V_z$. In Fig.~\ref{fig_Gamma_semi_fixed1island}(a), we vary $V_{z,c}<V_z<2$ meV for three different $\Delta =0.1, 0.2, 0.5$ meV, while fixing $\alpha=0.2$ eV \AA,  and $\mu=0$. This simulates the varying magnetic field continuously in experiments.  The trend of $\Gamma(V_z)$ is well captured by the analytical results in Eqs.~(\ref{eq:kF_semi_wire_1}), (\ref{eq:kappa_semi_wire_1}) and (\ref{eq:Gamma_1island}), from $V_z$ slightly larger than $V_{z,c}$ up until the decay length approaches the island length, $1/\kappa < l$. The fast oscillation is through $\cos^2(k_F l)$ and the slower envelope is due to $\kappa e^{-\kappa l}$.  After that $\Gamma(V_z)$ saturates with increasing $V_z$ instead of  going to zero ($\propto\kappa$), as the Majorana overlap (as well as the net charge) maximizes.  As discussed before, different $\Delta$ changes $\kappa$ mostly by $\kappa\propto \Delta$, in addition to a shift $V_z\rightarrow V_z-\Delta$. Figure~\ref{fig_Gamma_semi_fixed1island}(b) shows similarly $\Gamma(V_z)$ for three $\alpha=0.1, 0.2, 0.5$ eV\AA, while $\Delta$ is fixed at 0.2 meV. It shows the same role played by $\alpha$ and $\Delta$ in affecting $\kappa$, except that $\alpha$ does not shift $V_z$. We can compare Figs.~\ref{fig_Gamma_semi_fixed1island} (a) and (b) with Figs.~\ref{fig_Gamma_N100_q0l1} (a) and (b), where $\mu$ affects $k_F$ the same way as $V_z$ here but does not affect $\kappa$ away from the topological transition.

Next we show the effect of tuning the nanowire chemical potential $\mu$ on $\Gamma$ in Figs.~\ref{fig_Gamma_semi_fixed1island} (c) and (d). (c) and (d) follow (a) and (b) in their parameter choices except for fixing $V_z$ to be a typical 0.7 meV and varying $\mu$. As one could see, the major effect of varying $\mu$ is the oscillation of $\Gamma(\mu)$, which is through the changing $k_F(\mu)$, with a slowly-changing oscillation amplitude  reflecting the weak dependence of $\kappa(\mu)$ as we have discussed. At larger $\Delta$ or $\alpha$, there is a more perceivable change of the oscillation amplitude as a function of $\mu$, also mentioned before. Near the topological transition boundary, which is not of practical interest for Majorana qubits, $\Gamma$ is often enhanced (depending on the $\mu$ value) from the analytical estimation due to a very slow decay component $\propto e^{-wx}$, $w\rightarrow0$ \cite{DasSarma_PRB12,BenShach_PRB15}. The similarities between the $\mu$ dependence in the semiconductor model and the Kitaev model can be seen comparing Figs.~\ref{fig_Gamma_semi_fixed1island} (c) and (d) with Figs.~\ref{fig_Gamma_N100_q0l1} (a) and (b) where $\mu$'s main effect is through $k_F(\mu)$.

To be complete, we also show the complementary $\Gamma(\Delta)$ and $\Gamma(\alpha)$ in Figs.~\ref{fig_Gamma_semi_fixed1island} (e) and (f), though it is not practical to tune them continuously in experiments. Again, the dependence is captured by the analytical expression when the decay length is smaller than the wire length. When $\kappa\ll 1/l $, however, instead of $\propto \kappa e^{-\kappa l} \rightarrow 0$, $\Gamma$ saturates.

\begin{figure}[!htbp]
\includegraphics[width=8.5cm]{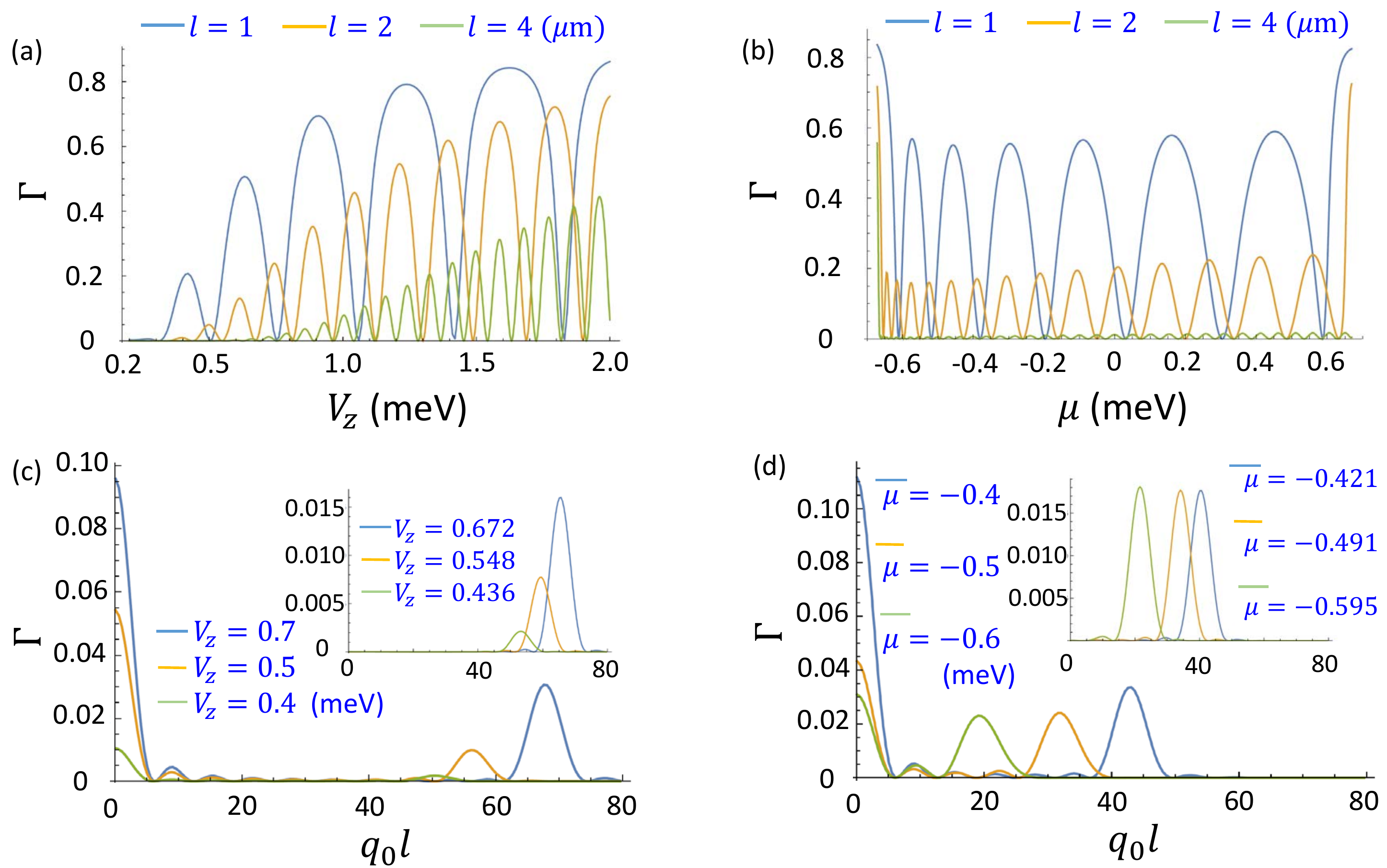}
\caption { $\Gamma$ with different island lengths $l$ (assuming one island dominates the results) and phonon wavevector $q_0$, with fixed $m^*=0.014 m_e$, $\Delta=0.2$ meV and $\alpha=0.2$  eV \AA. (a) $\Gamma(V_z)$ and (b) $\Gamma(\mu)$ for three different island lengths $l=1, 2$ and 4 $\mu$m with $q_0 l=1$. With $l=2$ $\mu$m fixed, $\Gamma(q_0 l)$ for (c) three different $V_z=0.7, 0.5, 0.4$ meV and fixed $\mu=0$,  and (d) three different $\mu=-0.6, -0.5, -0.4$ meV and fixed $V_z=0.7$ meV. In the insets of (c) and (d) we show fine tuned cases near $\cos(k_F l)=0$, such that $\Gamma(q_0l)$ is enhanced orders of magnitude around $q_0=2 k_F$ from the long phonon wavelength limit ($q\rightarrow 0$).
}\label{fig_Gamma_semi_l_q0}
\end{figure}

Now we consider the effect of the island length $l$ and the phonon wavevector $q_0$ while still assuming that one island dominates the result. First we vary  $l$ while keeping the product $q_0 l=1$ fixed. Its effects are twofold. Increasing $l$ decreases $\Gamma$ by reducing the Majorana charge ($\propto e^{-\kappa l}$), and increases the oscillation frequency in terms of  $V_z$ or $\mu$ $(\propto \cos(k_F l))$. They can be clearly seen in Figs.~\ref{fig_Gamma_semi_l_q0} (a) and (b) with three different $l=1, 2$ and 4 $\mu$m.

We focus on the effect of $q_0$ in Figs.~\ref{fig_Gamma_semi_l_q0} (c) and (d) to analyze the contribution from different phonon wavevectors. $q_0$ is eventually determined by the qubit energy difference in the system ($q_0=\varepsilon_{q_0}/\hbar v_l$). As shown in the analytical expression Eq.~(\ref{eq:Gamma_1island}), when $q_0\ll k_F$, $\Gamma(q_0)$ is proportional to $q_0^{-2} \sin^2(q_0 l/2)$. When $q_0$ approaches $2k_F$,  a near resonance in $\Gamma(q_0)$ appears as the integration over the real space results in a near delta function. This enhancement of $\Gamma$ is especially pronounced when the total Majorana charge is around zero, i.e., reaching the valleys of $|\cos(k_F l)|$ [Eq.~(\ref{eq:charge})]. At these spots, the increase of $\Gamma$ can be orders of magnitude when $q_0$ matches $2k_F$ [see insets of Figs.~\ref{fig_Gamma_semi_l_q0} (c) and (d)]. As the relevant phonon mode tends to be a long wavelength acoustic mode, the near resonances occur for relatively small $k_F$'s. That is when $V_z$ is tuned relatively small or when $\mu$ is tuned negative toward $-|V_z|$ [Eq.~(\ref{eq:kF_semi_wire_1})]. Once we are in the small $k_F$ region, we can fine tune to reach the vanishing total charge spots. Due to the insensitivity of $\kappa$ on $\mu$, tuning $\mu$ does not affect the resonance peak of $\Gamma$ [inset of Fig.~\ref{fig_Gamma_semi_l_q0} (d)] while the decreasing $V_z$ decreases $\Gamma$ peak somewhat at the same time [inset of Fig.~\ref{fig_Gamma_semi_l_q0} (c)]; also see Eq.~(\ref{eq:kappa_semi_wire_1}).  These nanowire results should be compared with Figs.~\ref{fig_Gamma_mu199-150_Delta001-1} (e) and (f) for the Kitaev model.

Trends of the barrier length ($l_B$) dependence are identical to those for the Kitaev model, as shown in Fig.~\ref{fig_Gamma_N100_mu198-199_barrier01-02-04} and in the associated discussions, which we do not repeat here.

Finally, we estimate the materials prefactor in front of the unitless quantity $\Gamma$ using InSb as the chosen material, while neglecting the contribution from the superconductor to these materials parameters. $\rho_l=\rho\times A=5.8 \times 10^3$ kg/m$^3 (5\times {10^{-8}})^2$ m$^2=1.45\times 10^{-11}$ kg/m (the nanowire cross sectional area $A$ is chosen to give a concrete estimate, while keeping in mind that $\Gamma\propto A^{-1}$ in the 1D limit), $\epsilon=16.8\epsilon_0$ ($\epsilon_0$ is vacuum permittivity), the averaged $\Xi_l, e_{m,l}$ and $v_l$ are taken to be $-7.3$ eV \cite{Vurgaftman_JAP01}, $7.1\times 10^{-2}$ C/m$^2$ \cite{Madelung_book04}, and  $3.7\times 10^3$ m/s corresponding to InSb acoustic phonon modes. As a result, numerically we have for the deformation potential related contribution,  and for the piezoelectric related contribution, respectively as,
\begin{eqnarray}
\tau^{-1}_{1D}&=&\tau^{-1}_{1D,{\rm dp}}+\tau^{-1}_{1D,{\rm pe}} \nonumber\\
&=&\frac{27}{{\rm meV}\; {\rm ns}} \frac{|T|^2}{\varepsilon_{q_0}} \Gamma +\frac{0.68 \;{\rm meV}}{ {\rm ns}} \frac{|T|^2}{\varepsilon_{q_0}^3} \Gamma. \label{eq:rate_numerical}
\end{eqnarray}
As mentioned before, $|T|\ll \varepsilon_{q_0}$ for  our perturbative formulation to apply. We can see  that in general, the smaller the phonon energy $\varepsilon_{q_0}$ is, and the larger the tunnel coupling $T$ is, the larger the relaxation rate $\tau^{-1}_{1D}$. The tunneling is estimated roughly to be  $T\approx m^*\alpha\Delta e^{-\kappa_U l_B}/(\hbar^2\kappa_U) $ between two semiconductor islands with the same $\Delta$, $\alpha$ and  $m^*$, and a tunnel barrier of height $U$ and width $l_B$, in the limit of large Zeeman energy $V_z\gg \{m^*(\alpha/\hbar)^2, \Delta\}$ and $U\gg V_z$ \cite{Aasen_PRX16}. Basically this result can be derived by matching the Majorana wavefunction in the island side $u_r(x)=e^{\kappa x}(A e^{i k_F x}+ B  e^{-i k_F x})$ and the wavefunction in the barrier $C e^{-\kappa_U x}$ as well as their derivative at the boundary $x=0$. We get $\frac{B}{A}\approx -\frac{2i k_F}{\kappa_U}-1$ and $\frac{C}{A}\approx -\frac{2i k_F}{\kappa_U}$. Normalizing the wavefunction, we have $A\approx \sqrt{\kappa}$. Then one uses the classic result $T\sim \kappa_U e^{-\kappa_U l_B} |C|^2$ \cite{Bardeen_PRL61,Sau_PRB11}. With this approach one can also calculate $T$  in more general parameter regimes.

\section{Summary}\label{sec:summary}

We have systematically studied the phonon-induced Majorana qubit relaxation in the tunnel coupled two-island setup (using two different Majorana models---Kitaev chain and nanowire), whose understanding is relevant  for implementing various topological  gate operations and quantum memory. The relaxation behavior is summarized below.

Apart from the materials parameters and the power-law relation with tunnel coupling and phonon energy, the relaxation rate essentially follows the dimensionless quantity $\Gamma$ [see Eqs.~(\ref{eq:rate_1D}) and (\ref{eq:rate_numerical})].  First, $\Gamma$ scales roughly with the Majorana charge squared which is basically controlled by $(\kappa e^{-\kappa l})^2$, with the decay length $1/\kappa$ and the island length $l$ as well as the oscillating function $\cos(k_F l)$ (eventually as a function of $V_z$, $\Delta$, $\alpha$, $l$, and  $\mu$). The effective charge oscillates around zero with a sensitive factor $\cos^2(k_F l)$, which could make the experimental measurement of the relaxation rate $\tau^{-1}$ appear rather randomly distributed, but the average magnitude can be  meaningfully compared with the theoretical calculation.
On top of that, we have the decaying dependence of $\Gamma$ on phonon wavevector $q_0$ at small $q_0$, the resonance behavior of $\Gamma(q_0)$, and the dependence of $\Gamma$  on tunnel coupling and phonon energy. The resonance occurs due to the matching of the oscillation of charge density and phonon mode in the real space (the phonon wavevector $q_0$ equals twice the Fermi wavevector $2k_F$).  The phonon coupling is the  strongest at these resonances when the Majorana total charge nearly vanishes. On the other hand, when the charge distribution is mostly of the same sign ($\cos^2(k_F l)\approx 1$), the strongest coupling with phonon happens for $q_0\rightarrow0$. The vanishing Majorana charge condition also alleviates the relaxation cancellation between the two islands.   Lastly and importantly,  we remark that $\varepsilon_{q_0}$ is not constrained by the Majorana splitting in the islands ($\varepsilon_{{\rm Maj},L/R}\equiv \varepsilon_{\rm o_{L/R}}- \varepsilon_{\rm e_{L/R}}$), but $\varepsilon_{q_0} = \varepsilon_{{\rm Maj},L}-\varepsilon_{{\rm Maj},R}\mp ( eV^{\rm lead}_L-eV^{\rm lead}_R)$, where $V^{\rm lead}_L-V^{\rm lead}_R$ is the external bias between the left and right lead electrodes (not to be confused with $\mu$ in each island, which is the relative chemical potential from the band bottom in the absence of $V_z$ and $\Delta$; the absolute chemical potential is always the same as that in the lead. See Fig.~\ref{fig_schematic}).  So there is a relatively easy experimental tuning parameter to control the relaxation. This important physics could be easily overlooked if one forgets about the relativeness of zero energy references in the TSCs and the physical significance of their difference. We do not include the change of superconducting phases which should not affect our conclusions about the effective charge distribution $|u_0(\mathbf{r})|^2-|v_0(\mathbf{r})|^2$.

From these results, we see that in order to achieve as slow a relaxation  as possible due to interaction with phonon, one can either have large $\kappa l$ (or fine tune the $\cos^2(k_F l)$ factor to vanish),  large $\varepsilon_{q_0}$, or small tunneling $T$, or preferably a combination of all of these factors. In operations which rely on tunneling, $T$ cannot be too small. The ability to change $V^{\rm lead}_L-V^{\rm lead}_R$ could readily reduce the relaxation, while not interfering with other parameters required for  qubit operations. We believe that tuning the lead potentials is perhaps the most effective way to suppress phonon-induced Majorana qubit relaxation.

Finally, it is illuminating to emphasize the similarity and difference between the relaxation of the  double-island MZM qubit and that of the double-quantum-dot charge qubit which has been realized in semiconductor quantum dots \cite{Hayashi_PRL03, Petta_PRL04, Fujisawa_science98} and resembles the case of very short wires. As mentioned in the Introduction, both types of qubits are encoded by the change of fermion occupation in the islands/dots, and the relaxation can be induced by the tunneling between the two islands/dots. While the electron tunneling between the quantum dots is assisted by the electric field perturbation (i.e., electron-phonon interaction or charge noise) \cite{Fujisawa_science98, Hayashi_PRL03, Fedichkin_PRA04,Vorojtsov_PRB05, Wu_PRB05, Stavrou_PRB05}, ideal MZMs are expected to be immune from this perturbation for its neutral charge distribution in the topologically protected infinite MZM separation limit. However, as it is well known by now, the charge distribution is far from zero for the possible MZM in experimentally available nanowires due to the wire length being not much longer than the MZM decay length or impurities effectively shortening the wire. This situation makes MZM qubit somewhat similar to the charge qubit, albeit a less efficient one with the effective charge being less than one electron. It also makes the MZM qubit susceptible to the same perturbation effect, perhaps to a lesser extent, as in the charge qubit. The distinct features of the MZM qubit relaxation come from the unique charge distribution of the electron-hole hybrid, the $|u_{i,0}|^2-|v_{i,0}|^2$ term displayed in Eq.~(\ref{eq:Gamma}). This work has been largely devoted to analyzing these features. Aside from the total charge being small, this electron-hole hybrid distribution varies as a function of chemical potential, Zeeman energy, superconducting pairing and spin-orbit coupling,  and has additional dependence on phonon wavelength coming from the 1D nature of the wire (as opposed to the quantum dot), which do not arise in the regular coupled quantum dots situation.  The larger size of the island compared to that of the quantum dot also reduces the energy splitting due to charging energy, which we do not include in the current study. A similar interplay between the barrier length and phonon wavelength has been seen in double quantum dots \cite{Brandes_PRL99, Brandes_PRB02}. Lastly, the energy of the relevant phonon in the charge qubit also depends on the source-drain bias, with the small MZM spliting $\varepsilon_Q$ term [Eq.~(\ref{eq:E_q0})] replaced by the large charging energies in the dots \cite{Fujisawa_science98,  Hayashi_PRL03}.

Thus, the two-island four-MZM Majorana qubit has many similarities with a coupled double-quantum-dot charge qubit. It is not \textit{a priori} clear that the relaxation time in the Majorana qubit is much longer than the $\sim 10$ ns relaxation time typically found in charge qubits, despite the much smaller effective charge in the former. One may roughly expect that the charging energy in the quantum dot or island decreases with the effective charge squared (and thus with $\Gamma$ on average). The MZM energy splitting $\varepsilon_Q$ may decrease even faster than the charging energy when it is reduced below the Josephson coupling \cite{Aasen_PRX16}. In the situation without the external bias, $V^{\rm lead}_L-V^{\rm lead}_R=0$, it is likely that the piezoelectric contribution for the relaxation rate in Eq.~(\ref{eq:rate_numerical}) dominates over the deformation potential one (assuming the MZM energy splitting $\varepsilon_Q \ll \sqrt{27/0.68}\approx 6$ meV). Then the decrease of the $\Gamma$ quantity in the numerator of Eq.~(\ref{eq:rate_numerical}) is outpaced by the $\varepsilon_Q^3$ term in the denominator, and as a result $\tau^{-1}_{1D}$ can be far greater for the Majorana qubit than for the charge qubit from their interaction with phonons and puts an upper limit for the Majorana qubit lifetime.
The overall relaxation has contributions also from charge noise and other sources, which are difficult to estimate without specific device and noise parameters.

The situation is different however, when one allows for the external bias. This term could readily dominate over the $\varepsilon_Q$ term for $\varepsilon_{q_0}$ in Eq.~(\ref{eq:E_q0}) and considerably suppress the Majorana qubit relaxation induced by phonons, and also likely by the low-frequency charge noise. At the same time, the external bias does not affect much the charge qubit relaxation except for near resonant tunneling, because the quantized energy level is already large in the quantum dots. One should also keep the above discussion in mind when dealing with Majorana wires that are much shorter than the Majorana decay length. We stress that our perturbative result only applies for the tunnel coupling $T\ll \varepsilon_{q_0}$ regime. So it excludes the scenario when the external bias is set to zero, the MZM splitting $\varepsilon_Q$ is tiny, and $T$ becomes larger than $\varepsilon_Q$ during active tunneling operations or due to noise on the barrier. In this regime, decreasing the MZM effective charge and splitting decreases the quantity $\Gamma$ without the power-law divergence of the  relaxation rate denominator, and leads to better MZM lifetimes. Therefore, shortening the wire length may initially increase the relaxation rate when $\varepsilon_Q< T$ before decreasing it as discussed above (with external bias set to be zero).

Our results derived in this paper are expected to apply to the current Majorana nanowire experimental situation where the MZM energy splitting is unlikely to be very small either because the wire length is typically comparable to the coherence length, or perhaps because the apparent zero energy states in one (or both) island(s) are accidental near-mid-gap Andreev bound states \cite{Liu_PRB17} rather than truly topologically protected Majorana states.  In such situations, our approximation of $T \ll \varepsilon_Q$ should be well valid generally.
Experimental observations of qubit oscillations (Rabi or Ramsey type phenomena) in the Majorana qubit need to be reported before more progress in understanding and perhaps suppressing qubit relaxation can be made.

\section*{Acknowledgement}
We thank Will Cole and Jay Sau for helpful discussions. This work is supported by Laboratory for Physical Sciences and Microsoft.

\appendix
\section{Additional numerical results of $\Gamma$}\label{app:more_result}

\begin{figure}[!htbp]
\includegraphics[width=8.5cm]{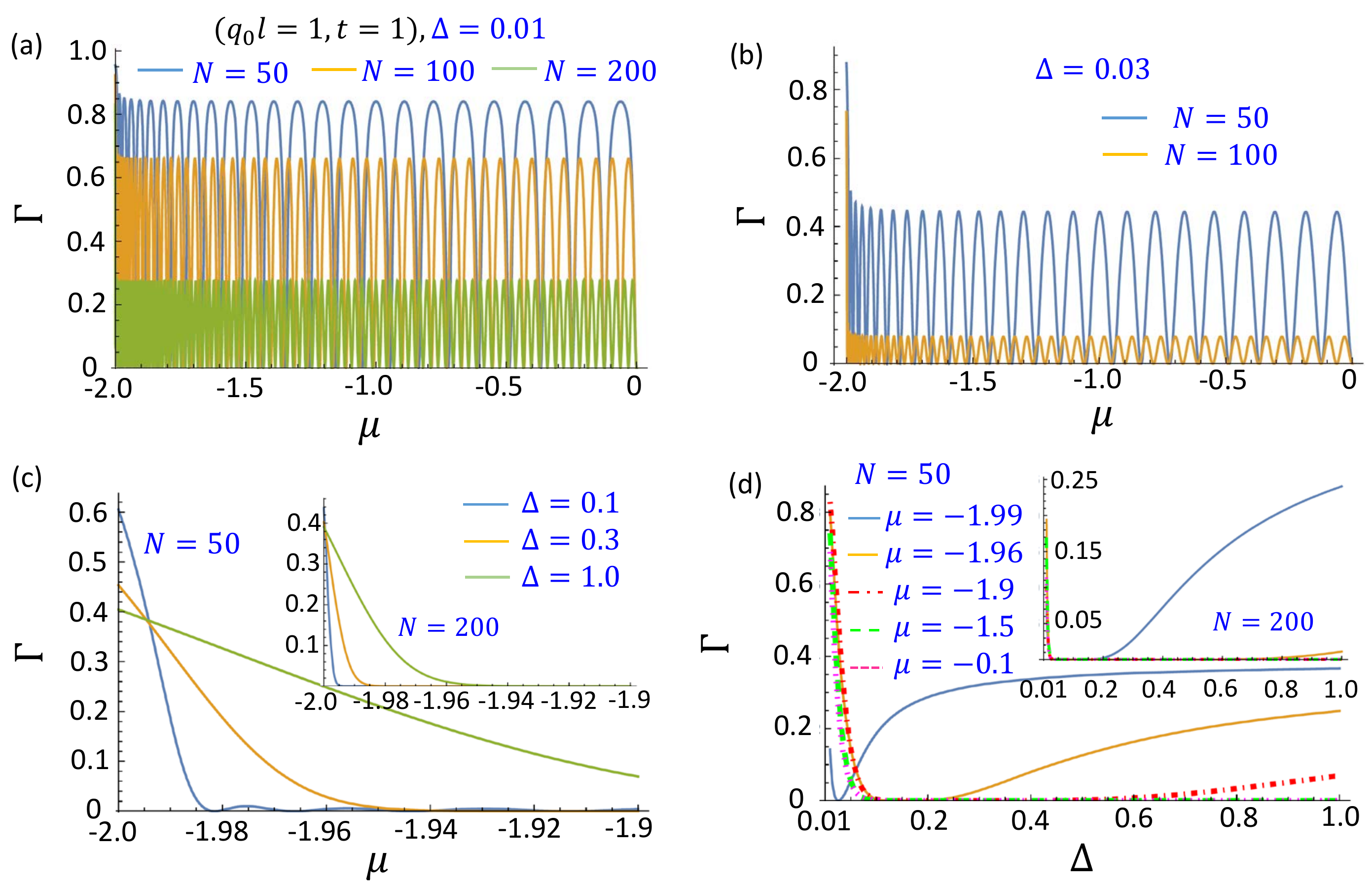}
\caption {Extension of Fig.~\ref{fig_Gamma_N100_q0l1} with $N$ changed from 100 to 50 or 200. In (b) the $N=200$ curve is too low to be visible and not shown.
}\label{fig_Gamma_N50-100-200_q0l1}
\end{figure}

\begin{figure}[!htbp]
\includegraphics[width=8.5cm]{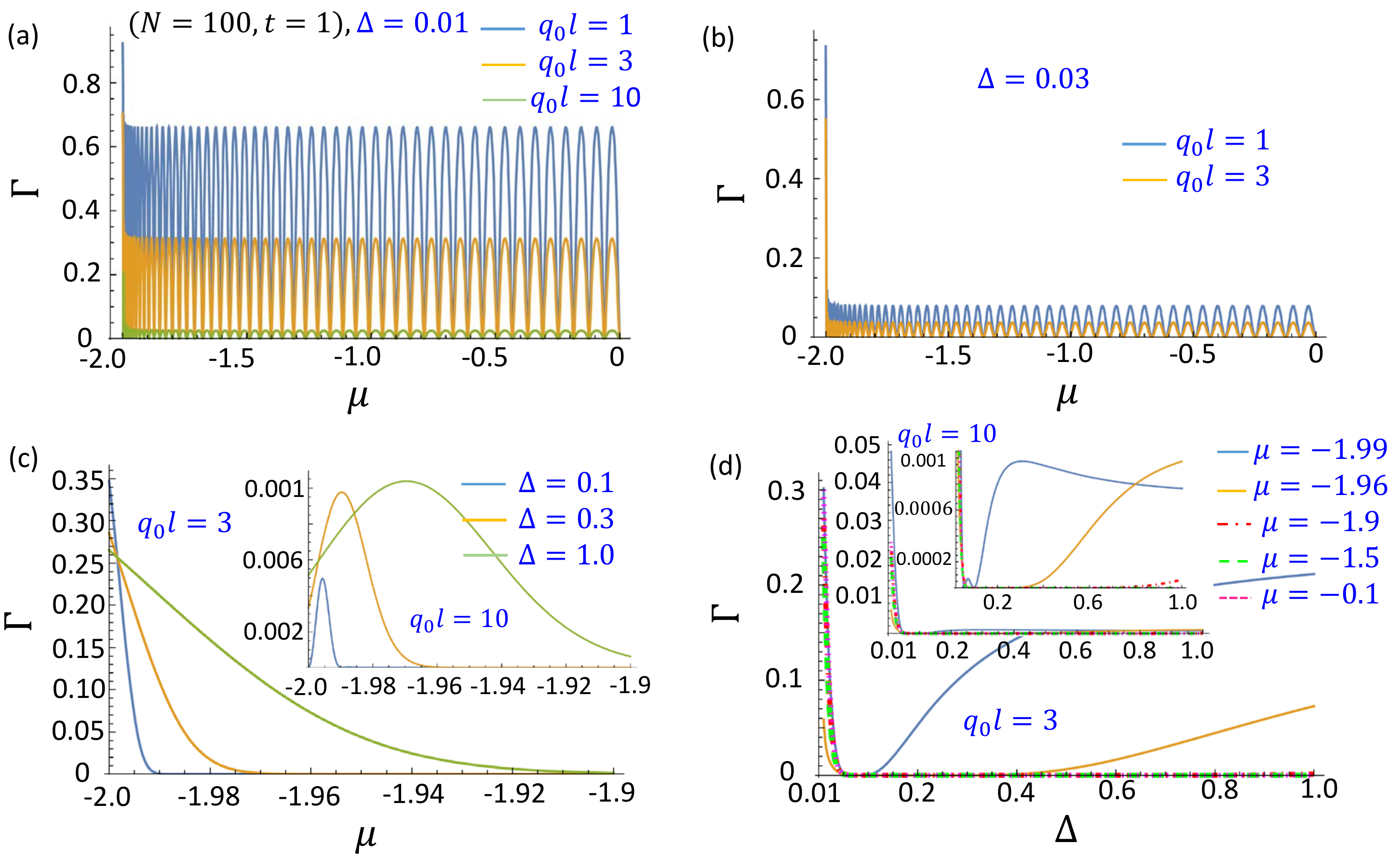}
\caption {Extension of Fig.~\ref{fig_Gamma_N100_q0l1} with $q_0l$ changed from 1 to 3 or 10. In (b) the $q_0 l=10$ curve is too low to be visible and not shown. In (d) the two insets are the same results shown in different scales, from which we can see for $q_0 l=10$ the similar saturation following the rise of $\Gamma (\Delta)$, albeit at much lower $\Gamma$ values than for smaller $q_0 l$'s.
}\label{fig_Gamma_N100_q0l1-3-10}
\end{figure}

In this appendix, we show additional results from Sec.~\ref{subsec:Kitaev_general_para}, the counterparts of Fig.~\ref{fig_Gamma_N100_q0l1} with $N=50$ or 200 in Fig.~\ref{fig_Gamma_N50-100-200_q0l1}, and  with $q_0l=3$ or 10 in Fig.~\ref{fig_Gamma_N100_q0l1-3-10}.

The general trends are that $\Gamma$ decreases with increasing $N$, that is, the island length, due to the smaller overlap, and $\Gamma$ decreases with increasing $q_0 l$ due to the faster oscillation of phonon field and thus larger cancellation of the $\Gamma$ integral. In Fig.~\ref{fig_Gamma_N100_q0l1-3-10} (c), $q_0 l=10$ is large enough for some visible resonance behavior near $\mu=-2$ and at large $\Delta$ values.

$\\$
$\\$
$\\$
$\\$
$\\$


\begin{thebibliography}{99}
\bibitem{footnote} The prefactor in this tunnel coupling contains the spin-orbit coupling $\alpha$, further reducing its magnitude (see reference \cite{Aasen_PRX16}). As a result, for a typical tunnel barrier height of 2 meV, the tunnel coupling $T$ is estimated to be the order of $10^{-4}$ meV \cite{Aasen_PRX16}, much smaller than most long wavelength acoustic phonon energy.
\end{thebibliography}
\end{document}